\documentclass{aa}
\usepackage{graphics} 
\usepackage{epsf} 
\begin{document}
\title{Low resolution spectroscopy  of ISOGAL sources:
Search for early-type stars with infrared excess 
\thanks{This is paper no.~9 in a refereed journal
based on data from the ISOGAL project}
\thanks{Based on observations with 
ISO, an ESA project with instruments funded by ESA Member States (especially the
PI countries: France, Germany, the Netherlands and the United Kingdom)
and with the participation of ISAS and NASA}
\thanks{Based on observations collected at the European Southern 
Observatory, La Silla Chile (63.L-0319)}}

\author{M.~Schultheis\inst{1}
\and M.~Parthasarathy\inst{2,3} 
\and A.~Omont\inst{1}
\and M.~Cohen\inst{4}
\and S.~Ganesh\inst{5}
\and F.~Sevre\inst{1}
\and G.~Simon\inst{6}  
}
\authorrunning{M. Schultheis et al.}
\titlerunning{Spectroscopy of ISOGAL sources}  
\offprints{schulthe@iap.fr} 
\institute{Institut d'Astrophysique de Paris, CNRS, 98bis Bd Arago, F-75014 Paris     
\and Indian Institute of Astrophysics, Koramangala, Bangalore, 560034, India 
\and National Astronomical Observatory of Japan, 2-21-1 Osawa, Mitaka, Tokyo 181-8588, Japan 
\and Radio Astronomy Laboratory, University of California, Berkeley, CA 94720, USA  
\and Physical Research Laboratory, Navarangpura, Ahmedabad 380009, India 
\and DASGAL CNRS UMR 8633, Observatoire de Paris, France            
}  
  
\voffset 1.0truecm

\date{Received August 8 2001  / Accepted .. .........}

\sloppy

\abstract{
An analysis of low resolution spectra and infrared data of 29 ISOGAL-DENIS sources with mid-IR excess
is presented. Eight ISOGAL sources from our sample with 7--15\,$\mu$m excesses are found to be
B and A-type stars, some of them with emission lines. Two ISOGAL sources, J175614.4-240831
 (B3-4IIIe) and J173845.3-312403 (B7IIIe), show a bump between 5000 and 6000\,\AA~ which may be
attributed to extended red emission (ERE). Some of the B,A and F-type giants with   
a large infrared excess might be in the post-AGB phase.\\
For about 50\% of the sources in this preliminary study, a nearby second (or even multiple) component
was found. Such sources, in particular two B-stars, are not discussed when the probability of
the optical spectrum being associated with the ISOGAL source is low.
These results confirm that the  DENIS-ISOGAL I-J/K--[15] diagram is the most suitable diagram to distinguish between early (AB) and late  spectral types (KM). It provides  the most useful tool
to  systematically search  for nearby early-type stars with an infrared excess
 among the background of distant AGB stars in ISOGAL fields of the Galactic disk.
} 

\maketitle

  
\section{Introduction} \label{introduction}  

	Infrared dust emission is one of the best  tools for
tracing circumstellar matter in various phases of stellar evolution: cocoons and disks in star formation, mass-loss (and circumstellar disks) in various
classes of evolved stars, Vega-type disks, etc. 

	Surveys with the ISO satellite, especially with the ISOCAM instrument (Cesarsky et al. \cite{Cesarsky96}), have opened new possibilities with 
a sensitivity typically two orders of magnitude better than  IRAS and an 
angular resolution ten times better. The ISOGAL 7 and 15$\mu$m survey 
(Omont et al. \cite{Omont2002}) has observed $\sim$16 deg$^2$ distributed 
in the most obscured regions of the central Galaxy. The total number of stars detected ($\sim 10^5$) is
comparable to the number of stars detected by IRAS in the entire Galaxy. ISOGAL and
DENIS (or 2MASS) near-infrared data constitute a  powerful combination for identifying
 the nature of ISOGAL sources even
in regions of high extinction (A$_{v}$ up to 20-30). The various colour-colour and colour-magnitude
diagrams from the photometry in the five ISOGAL-DENIS bands provide a wealth of information on the extinction, 
distance, intrinsic colours and absolute magnitudes. Most sources are
M giants at several kpc, with a few foreground K-giants, as observed by  
Ojha et al. (\cite{Ojha2000}). Among these M-type sources, a large proportion of those detected at 15~$\mu$m are AGB-LPVs with a variety of mass-loss rates which are well assessed
by the 15 $\mu$m excesses (Omont et al. 1999, Glass et al. 1999, Alard et al. 2001). 

	It is expected that this large sample of infrared sources contains a number of non-AGB stars with circumstellar material like Herbig Ae/Be stars, various types of post-AGB stars, and other peculiar
objects. Although the various colours and magnitudes offer broad diagnostics, spectroscopic information
is desirable to better characterise them and to identify their exact nature. 
We have thus begun a systematic spectroscopic study of peculiar ISOGAL sources, with the main goal to identify  accretion and post-accretion disks, post-AGB stars and Vega-type stars. Selection criteria are
designed to select hotter nearby stars with mid-IR excess while avoiding AGB
stars.
We report here the results of a preliminary study of low-resolution spectroscopy of a sample of 29 candidate objects.


\section{Observations}

Medium resolution spectra of 29 ISOGAL sources were obtained with the Danish 1.54\,m telescope at ESO, La Silla (Chile), equipped with the DFOSC instrument. The 
observations were carried out between 23--30 June 1999. 60\% of the time was 
lost due to bad weather.

The spectra were recorded on a Loral/Lesser 2052 x 2052 pixel CCD chip, and
cover a wavelength range from 3500\,$\rm \AA$ to 8200\,$\rm \AA$ with a resolution
 of  3\,$\rm \AA$/pixel. The objects were observed through a 2\arcsec~slit. Before each
spectrum an R-band image in the  was taken.
During the night a 10 sec exposure with a Ne-Ar lamp was made for calibration of the wavelength scale. At the beginning and at the end of 
each night one flux standard was observed (Feige 110 and Wolf 485a) 
through a 5\arcsec~slit. The observations were performed under 
photometric conditions.

\subsection{The sample}
 Due to the limiting sensitivity of the instrument, only sources 
with $\rm I \la 14$ were chosen.
 Such a condition excludes more than 90\% of ISOGAL sources, since
they suffer high extinction, as well as most AGB stars in the inner disk/bulge and 
most young stars with circumstellar and interstellar reddening. Therefore, with very few exceptions, the selected sources are expected to be foreground, relatively 
nearby ($\rm \la 1-2\,kpc$), objects.
We have chosen mainly sources with  large
values of $\rm K_{S}-[15]$ and $\rm [7]-[15]$ with respect to J--K and I--J. 
This implies the presence of circumstellar dust in a shell or disk.

 

Our objects were selected from  a prelimary version of the ISOGAL catalog but in the analysis reported here, we use the final ISOGAL catalog
(Omont et al. \cite{Omont2002}, Schuller et al. \cite{Schuller2002}). 
The DENIS observations form part of a 
dedicated survey of the Galactic Bulge (Simon et al., in preparation). The
2MASS data, when available, were taken from the second incremental data release (Skrutskie 
\cite{Skrutskie98}).

Table~\ref{Table1}  summarises the  ISOGAL sources observed, including
their DENIS, 2MASS and ISOGAL magnitudes as well as the projected distance in arcsecond between the apparently
associated DENIS and ISOGAL sources.  

\begin{table*}
\caption{ Coordinates (J2000), magnitudes (DENIS,ISOGAL and 2MASS) of the observed
sources as well as the projected distance (in arcsec) between the ISOGAL source and the
 DENIS counterpart, spectral type, E(B-V) derived by comparison with the Kurucz models. The last column specifies whether the object is discussed in
the main text (M) or has been rejected and therefore listed in the appendix (A)}
\begin{tabular}{lcllllllllllll}
\hline
\#&ISOGAL PJ&I&J&$\rm K_{S}$&[7]&[15]&$\rm J_{2MASS}$&$\rm H_{2MASS}$&$\rm K_{2MASS}$&dist.&Sp.type&E(B-V)&Flag\\
\hline
1 &143131.9-600646&14.0&8.7&$\rm 5.8^{*}$&5.6&4.3&--&--&--&0.5&F8-G0III&0.2&M\\
2 &143313.2-603705&11.5&10.0&9.0&5.2&4.5&--&--&--&0.9&K4III-II&0.2&A\\
3 &164825.8-441403&14.5&11.6&9.7&9.1&6.6&--&--&--&1.4&K4III-II&1.2&A\\
4 &165019.1-435927&14.7&12.3&9.6&7.2&6.6&--&--&--&1.7&K3III-V&1.0&A\\
5 &165102.4-443738&13.5&12.9&10.0&7.9&7.3&--&--&--&3.2&G4III&0.1&A\\
6 &172001.7-371024&11.4&10.7&10.2&9.6&7.9&--&--&--&0.4&G4III&0.1&A\\
7 &172807.6-331812&13.5&11.3&10.1&8.9&8.6&11.3&10.4&10.2&2.2&G8III-V&1.0&A\\
8 &172830.3-332214&10.7&9.5&8.5&7.8&7.7&9.5&8.7&9.2&2.0&K2III-II&0.1&A\\
9 &172823.5-351559&13.7&8.3&6.0&5.6&4.4&8.4&6.7&5.9&1.5&M5III&2.6&A\\
10&173219.0-334751&15.1&12.0&8.9&5.1&3.1&11.7&10.1&8.7&1.7&Be-Ae&2.5&A\\
11&173221.5-335405&13.4&9.3&6.9&6.5&6.3&9.2&7.5&6.9&0.2&M5III&2.0&A\\
12&173223.0-320841&12.2&9.8&8.3&7.7&7.2&9.8&8.6&8.2&1.1&K5III-II&0.7&A\\
13&173537.4-335753&14.3&11.9&9.8&7.3&4.8&11.8&10.5&9.7&0.6&lateF&1.5&A\\
14&173712.5-311852&11.0&10.2&9.7&8.7&7.8&10.2&9.8&9.7&0.7&B8IIIe&0.3&M\\
15&173745.0-311531&13.4&12.3&11.3&9.8&7.9&12.2&11.8&11.4&0.5&A0III&0.6&M\\
16&173845.2-312403&11.0&9.8&9.4&9.0&--&9.9&9.6&9.5&1.9&B7IIIe&0.5&M\\
17&174634.7-282008&11.6&9.8&8.7&7.8&6.3&9.8&8.9&8.7&2.5&K4III&0.3&M\\
18&174647.0-283542&12.1&10.5&9.3&8.0&7.1&10.5&9.7&9.3&1.5&K5III-II&0.2&M\\
19&174650.3-284451&13.1&12.1&10.9&8.5&--&12.2&11.7&10.9&1.4&B9III&0.7&A\\
20&174717.0-290144&11.1&8.6&7.0&6.4&6.2&8.6&7.4&6.9&2.1&lateG&1.7&A\\
21&174735.9-274633&13.4&10.7&9.6&8.3&7.3&--&--&--&1.7&M6III&0.2&A\\
22&174803.0-282019&12.6&11.9&11.3&8.0&7.1&11.9&11.5&11.1&1.6&G2III&0.1&M\\
23&174804.2-281028&13.2&10.6&9.0&9.2&--&10.6&9.5&8.9&0.4&K4III-II&1.5&A\\
24&174812.1-282319&11.0&8.4&6.7&6.4&6.3&8.3&7.1&6.6&0.9&M2III&0.2&A\\
25&174819.8-280513&14.1&12.1&10.0&7.7&7.5&--&--&--&0.9&K4III&1.0&A\\
26&175141.9-262754&12.8&10.0&8.8&7.7&4.0&10.1&9.3&8.8&1.5&early-type&3.2&M\\
27&175614.4-240831&11.2&10.4&9.9&9.0&8.3&10.3&10.1&9.8&1.0&B3-4IIIe&0.5&M\\
28&180328.4-222259&13.0&10.5&9.1&7.0&7.3&--&--&--&1.0&early-type&2.5&M\\
29&180446.3-300318&12.2&9.1&7.6&6.6&5.5&--&--&--&1.2&M7III&0.2&A\\ 
\hline
\end{tabular}
*  saturated in DENIS

\label{Table1}
\end{table*}

\section{Data reduction}

All reduction procedures were carried out with the MIDAS software package.
After removal of cosmic ray events, subtraction of the bias level
and the dark current, all CCD frames were divided by a normalized flat field.
Two sky frames were taken, left and right of the object and the mean sky was
subtracted from the object.
For the wavelength calibration, 12 Ne-Ar lines were used which led to a rms
error for each spectrum of about 0.2\,$\rm \AA$. The spectra were rebinned 
to a linear scale, with a resolution of 3\,$\rm \AA$/pixel. 
The two photometric standards Wolf 485a and Feige 110
(Stone \cite{Stone77}, Massey \& Strobel \cite{Massey88}) were used for flux calibration. All
spectra were calibrated with the averaged instrumental response from
the two standards and  were corrected for atmospheric extinction. 
 Typical errors in the absolute fluxes are  $\sim$ 10\%.


\section{Analysis}

\subsection{Spectral classification}

\begin{figure*}
\epsfxsize=7cm

\centerline {\epsfbox[20 130 710 590]{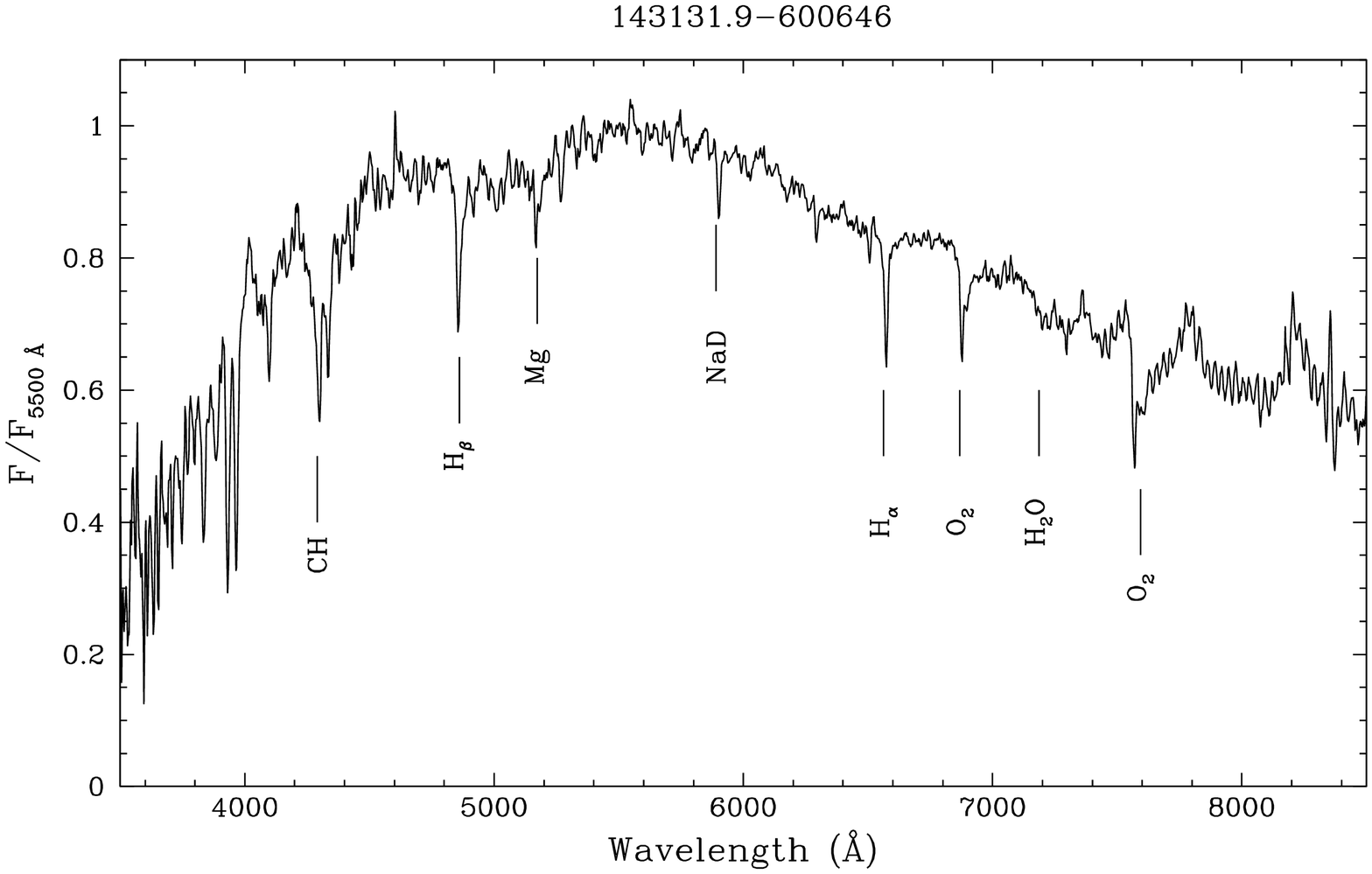} \epsfxsize=7cm
 \epsfbox[20 130 710 590]{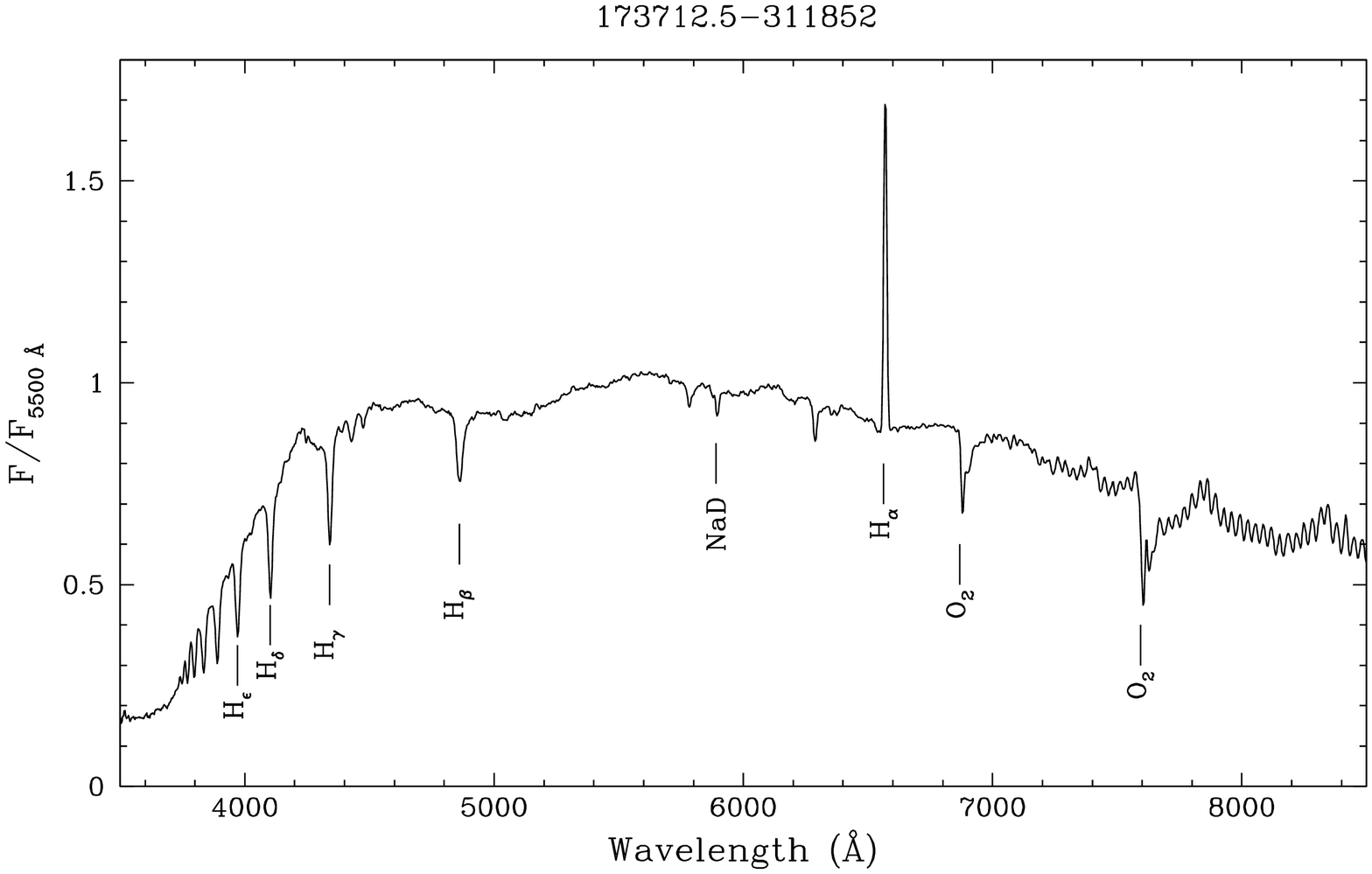}}

\epsfxsize=7cm
\centerline {\epsfbox[20 130 710 590]{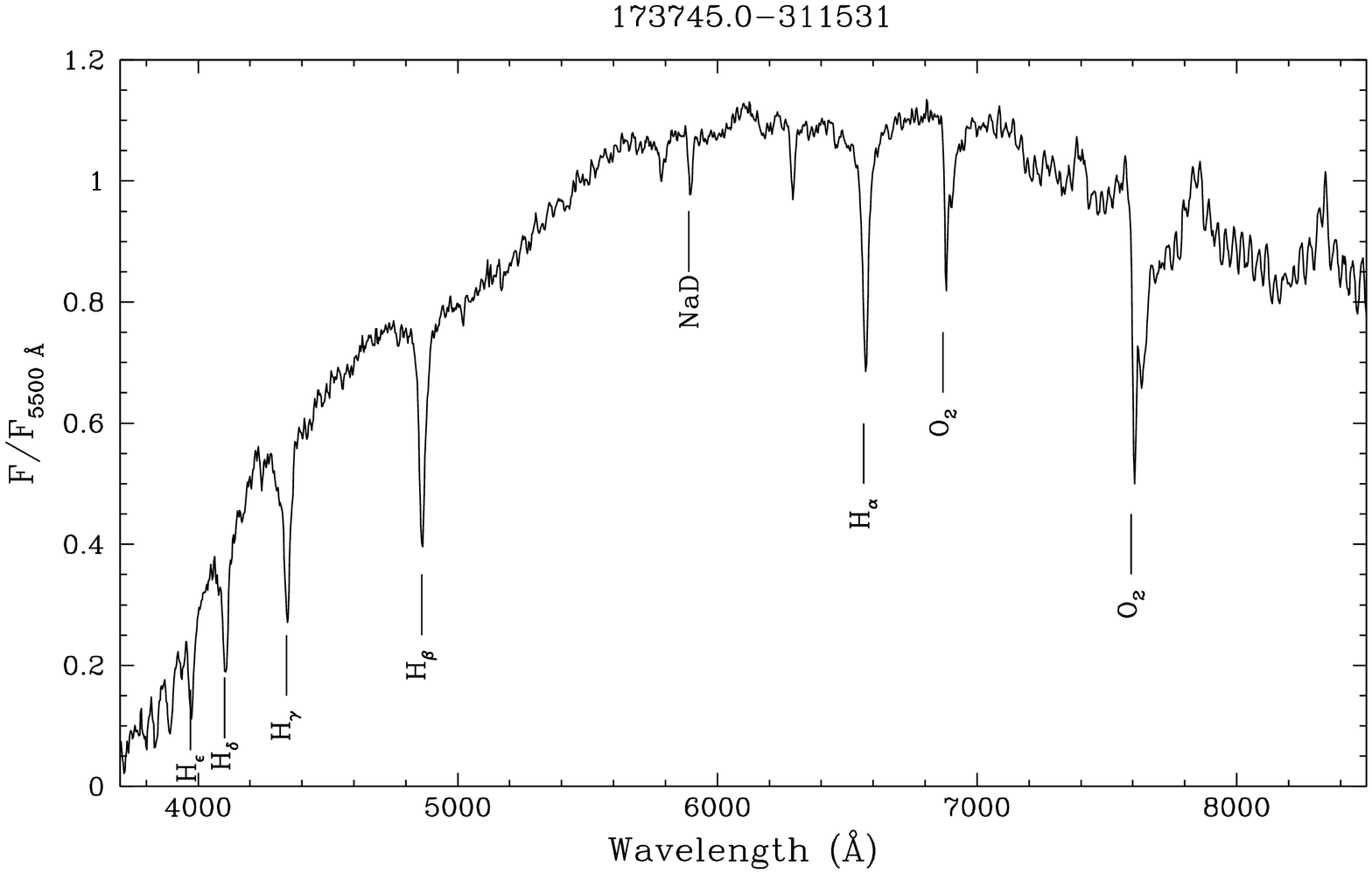} \epsfxsize=7cm
 \epsfbox[20 130 710 590]{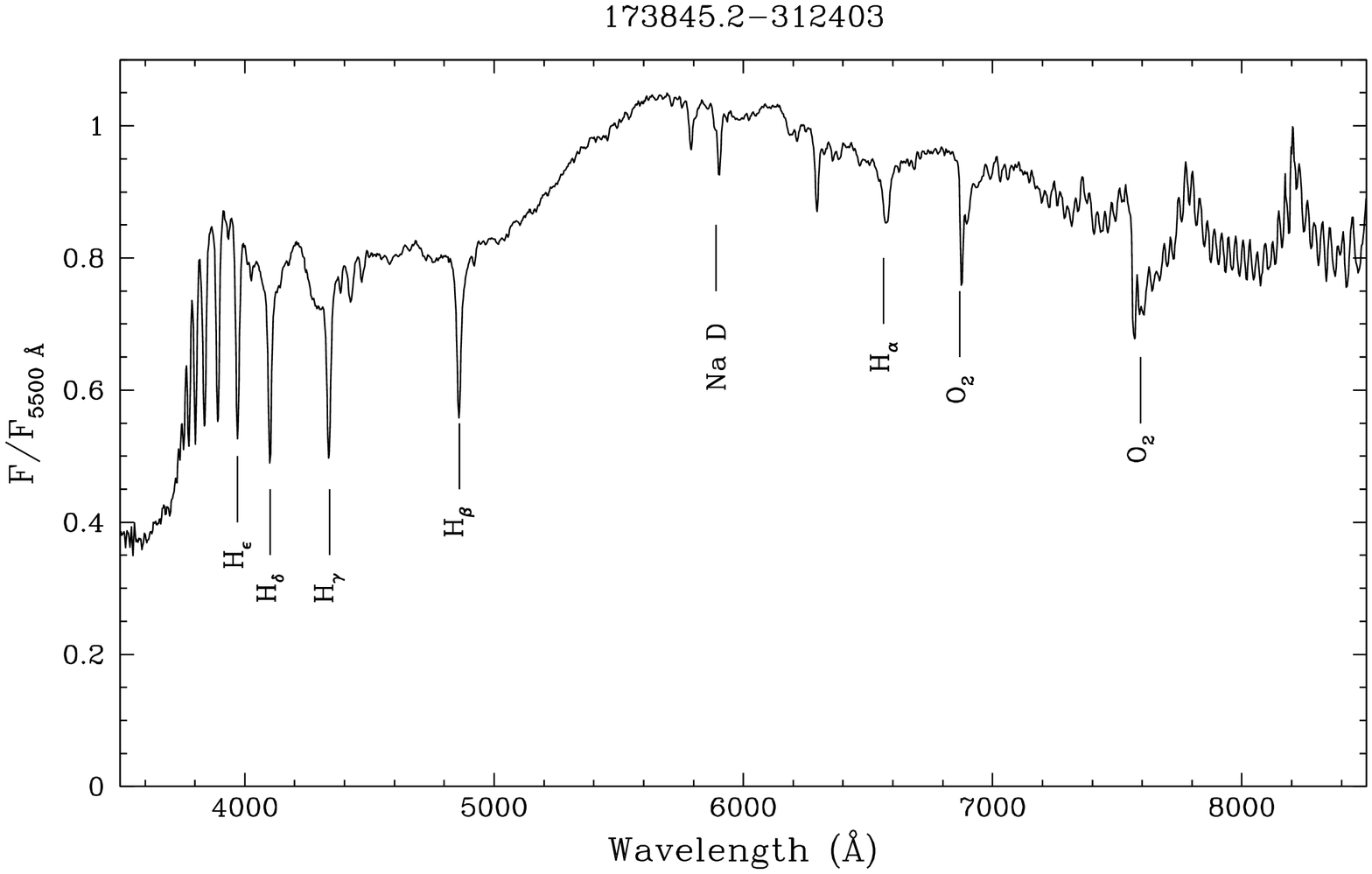}}

\epsfxsize=7cm
\centerline {\epsfbox[20 130 710 590]{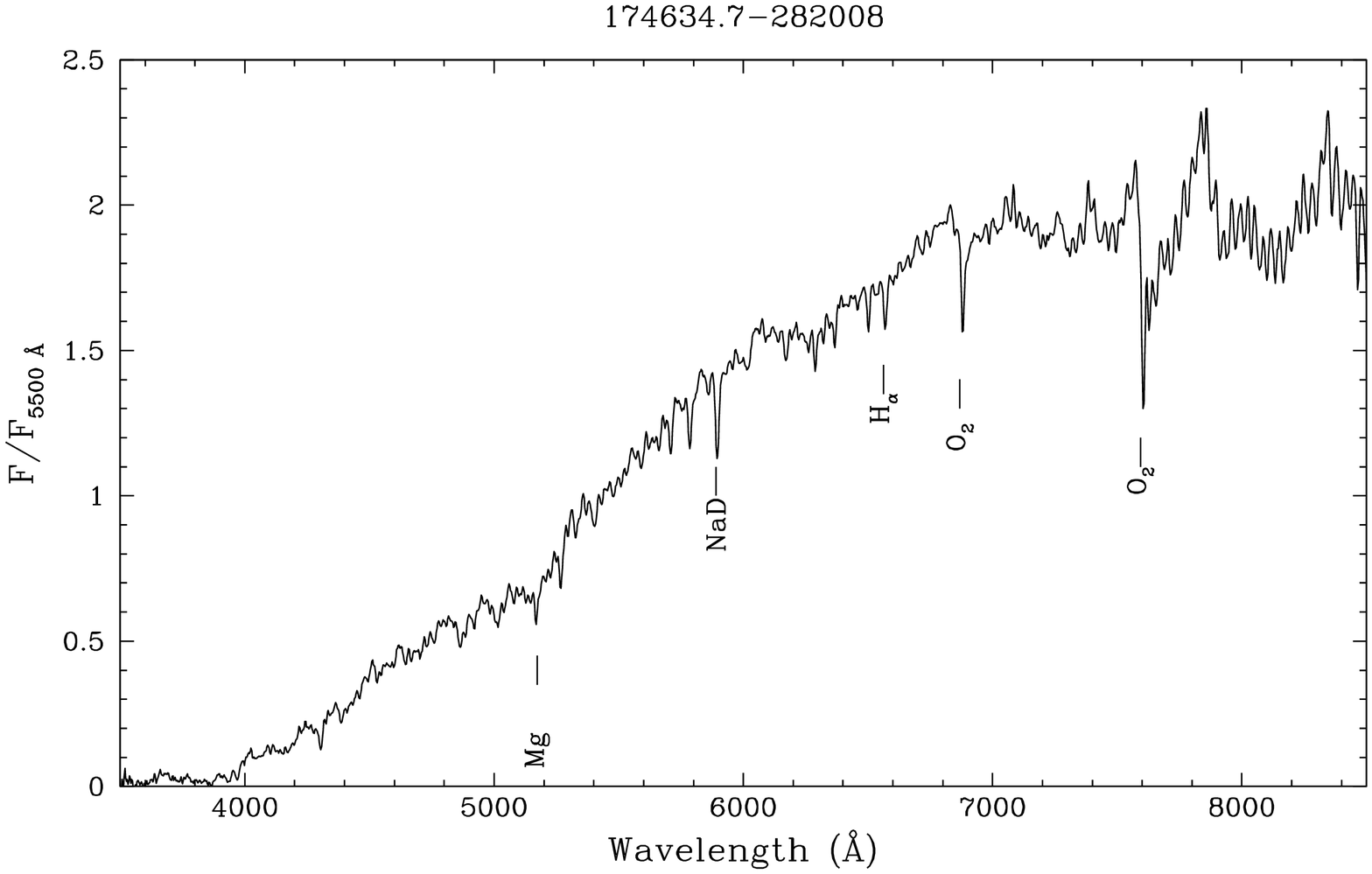} \epsfxsize=7cm
 \epsfbox[20 130 710 590]{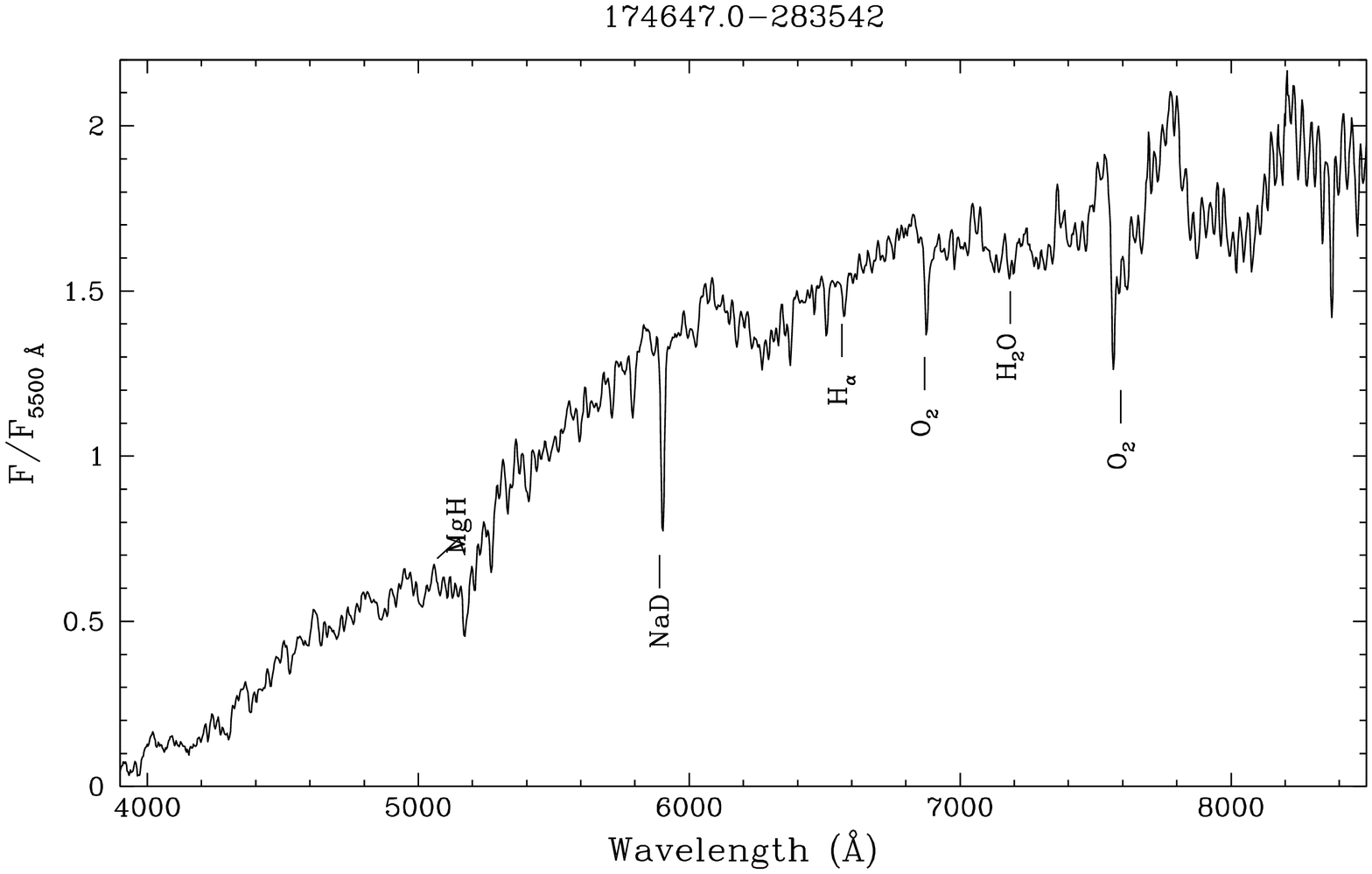}}

\epsfxsize=7cm
\centerline {\epsfbox[20 130 710 590]{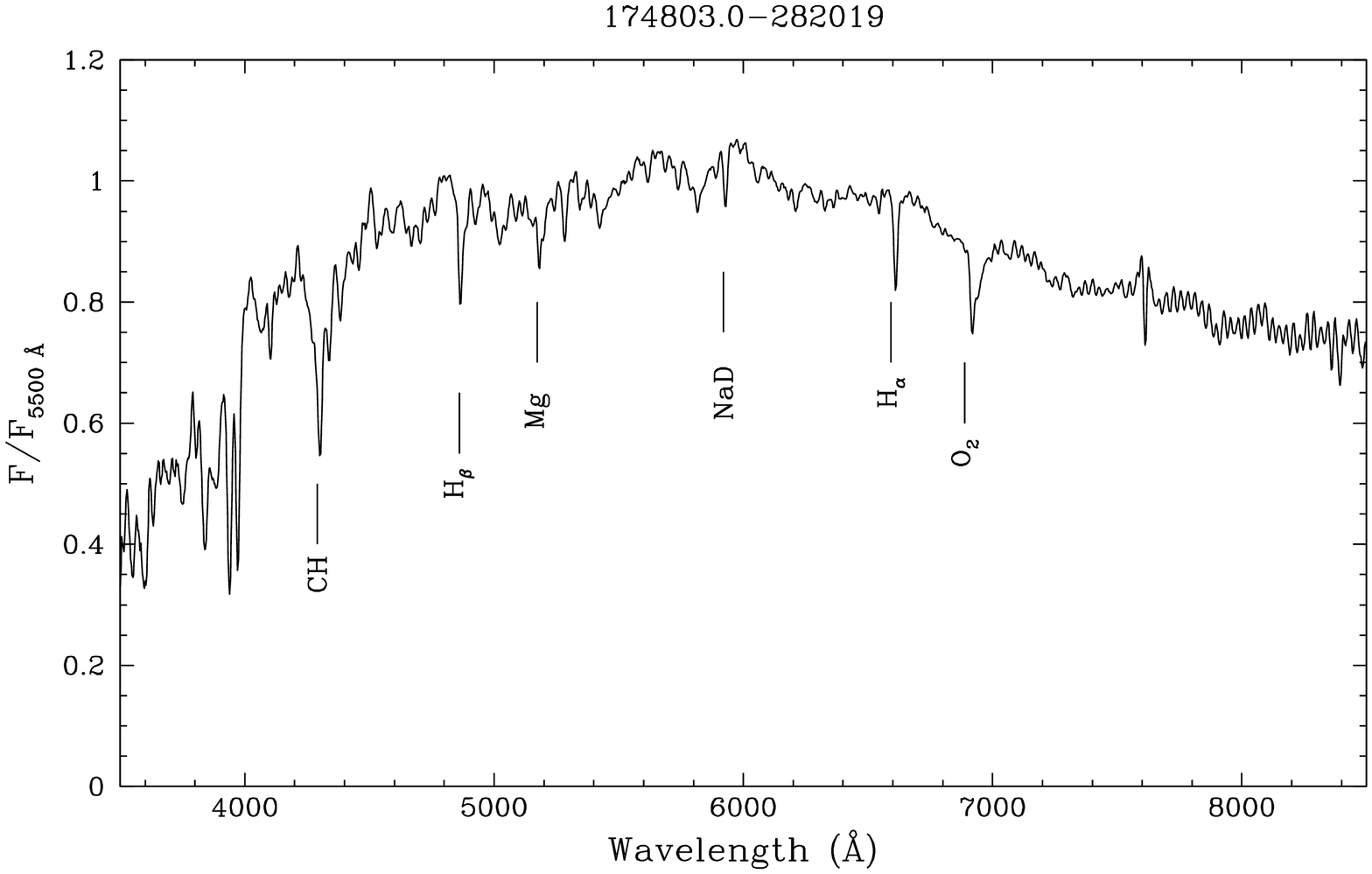} \epsfxsize=7cm
 \epsfbox[20 130 710 590]{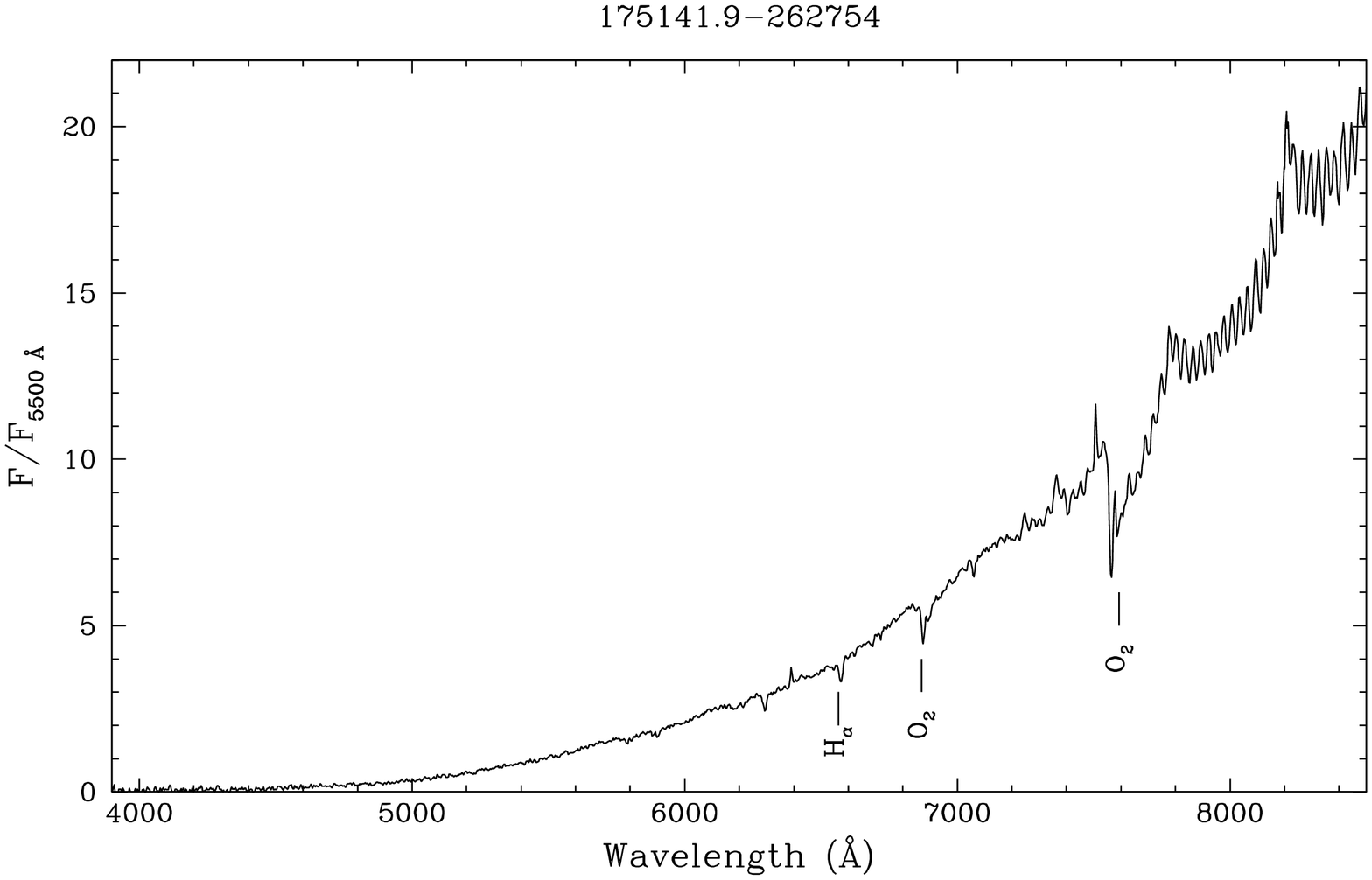}}

\epsfxsize=7cm
\centerline {\epsfbox[20 130 710 590]{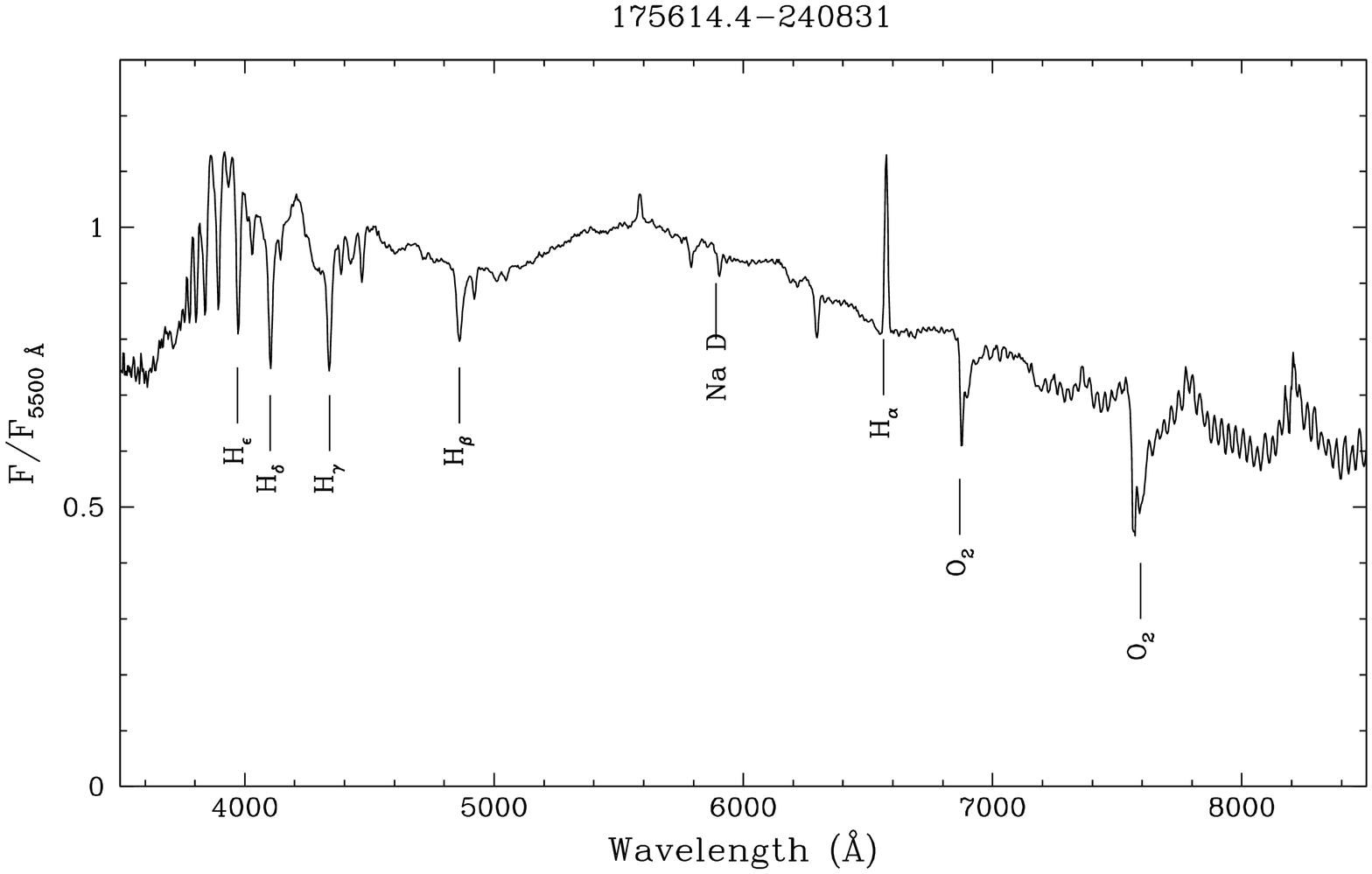}  \epsfxsize=7cm 
\epsfbox[20 130 710 590]{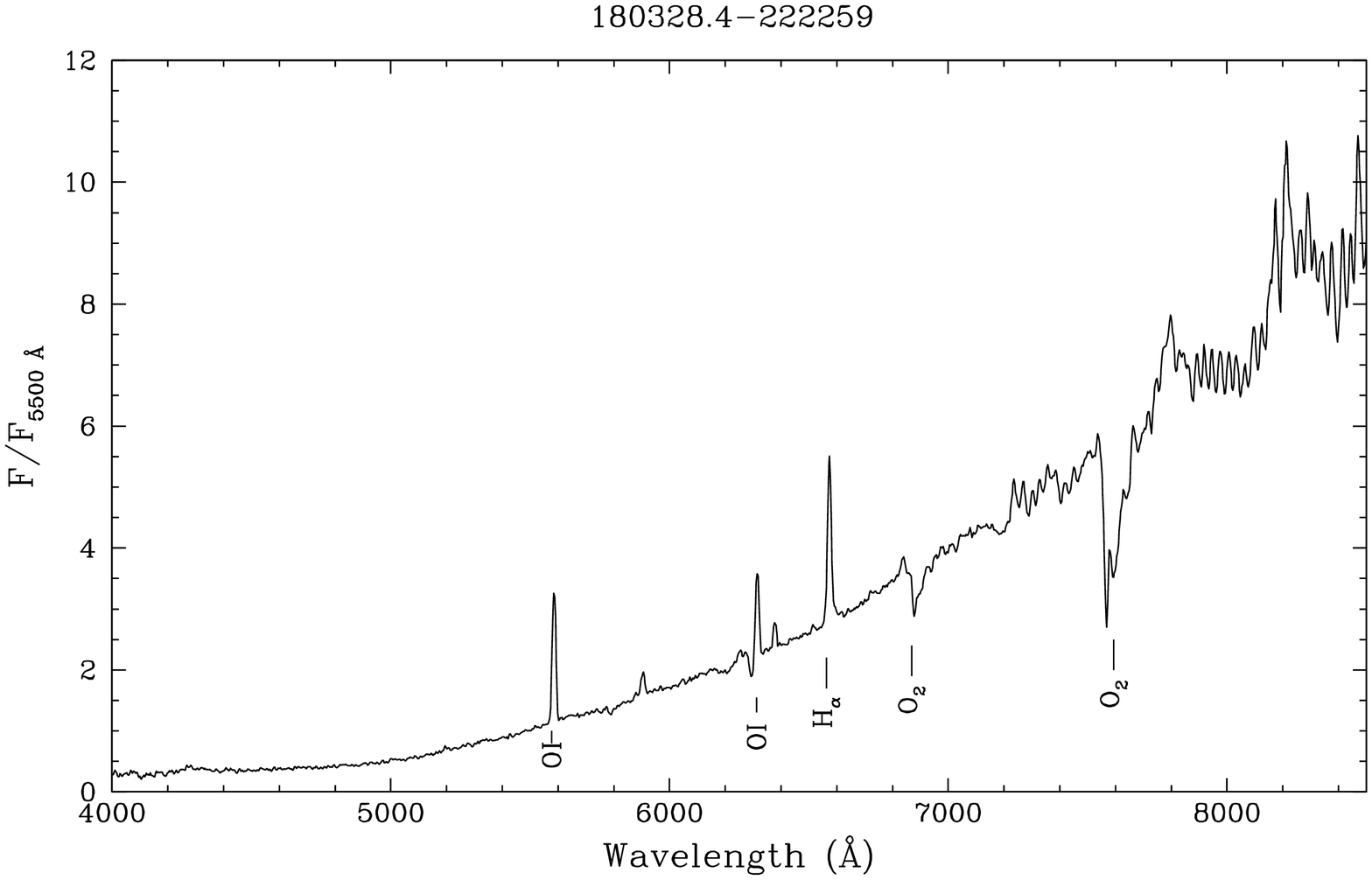}}

\caption{Sample of optical spectra of ISOGAL sources which are discussed
in the main text with the main line identifications. 
The spectra are normalized to unity at 5500\,$\rm \AA$.}
\label{spectra}
\end{figure*}

Figure \ref{spectra} shows a sample of the resulting spectra (F$_\lambda$ in erg s$^{-1}$ cm$^{-2}$ \AA$^{-1}$),
 normalized to unity by the flux
at 5500\,\AA. The most prominent  spectral features are indicated.
 A wide variety of spectral types is present.
We classified the optical spectra of our sources by comparing them
with the spectra of standard stars drawn from the low-resolution
spectral atlases of Jacoby et al. (\cite{Jacoby84}),  Torres-Dodgen \&  
Weaver (\cite{Torres93}), Pickles (\cite{Pickles98}),  Walborn (\cite{Walborn80}) and Yamashita et al. (\cite{Yamashita77}). 
The spectral types are given in Table 1. Some of our program 
stars show significant reddening and, therefore, the blue part of the 
spectrum shortward of 5000\AA~ does not have a high signal-to-noise
ratio. When the reddening is high, we have  used the 
red part of the spectrum of our program stars for classification.
The spectral atlases of Turnshek et al. (\cite{Turnshek85}), Walborn (\cite{Walborn80}) and
Torres-Dodgen \& Weaver (\cite{Torres93}) were useful in the red
spectral region for both hot and cool stars.
The uncertainty in our assigned spectral types is estimated to be two
subtypes, while (due to the low resolution of our spectra) the uncertainty
in the luminosity should be one subclass.

\subsection{Interstellar reddening}

We derived a rough effective temperature for each object using the
corresponding Kurucz models. The total (interstellar+circumstellar)
reddening has been obtained adopting the value which gave the best fit
between the observed spectrum and the reddened model spectrum (Fig.~\ref{SED}).
This method depends strongly on the adopted fundamental parameters ($\rm T_{eff}$, log\,g, log\,Z) of the Kurucz models and on the slope of the
observed spectrum.

Thus we estimate the uncertainty of E(B-V) for spectral types
B-F of the order of $\sim$ 0.3 while, for later spectral types(KM),
 we have larger errors ($\sim$ 0.5) as the Kurucz models do not reproduce
cool stellar spectra well. The corresponding E(B-V) values are given in Table 1.

\subsection{Rejection of dubious cases}
	
	Our aim was  to characterize the objects and to search for clear cases of mid-infrared excess in stars with spectral types no later than K. We will thus focus the subsequent analysis on such cases, while we report non-confirmed, the  most ambiguous cases, and M-stars in Appendix A, and in figures available electronically at CDS. Such rejected sources are relatively numerous ($\sim$50\%) because of the preliminary nature of our selection criteria and of the quality of data available at the time of the observations. The major cause of rejection (see Appendix A) is evidence of multiple sources in the field with a large probability of false association of a foreground visible star with a distant AGB star,
invisible at optical wavelengths. For each individual source, the DENIS/2MASS images in {\it{I, J, H and $K_{S}$}}  and the ISOCAM images in 7 and 15\,$\mu$m  were examined by eye in order to 
check the cross-identification of the ISOGAL source with the near-IR
counterpart. For about 50\% of the sources, a second object was found close by, or even multiple components  (see the description of the individual objects). When the projected distance to the second source is less than 3.5\arcsec (the search radius for ISOGAL-DENIS associations)  we have carefully considered  the possible association of the ISOGAL source  with the second DENIS source (though we took the optical spectrum of the primary candidate). 
In addition,  we used the $\rm (I-J)_{0}/ (J-K)_{0}$ diagram (see Fig.~3b) to decide if those objects follow the expected temperature sequence in the colour-colour diagram indicated by the straight lines in Fig.~3b. 

 The objects  with  a second nearby source ($\rm < 3.5\arcsec$) for which the position in the colour-colour diagram (see Fig.~3b) does not agree
 with the temperature sequence,  were
rejected from further discussion (see Appendix). 

	In addition, we have also rejected the five identified M-stars, and star \#23 where the present higher quality  ISOGAL fluxes do not confirm the existence of a mid-infrared excess.

\subsection{Notes on individual objects}

\begin{figure*}
\epsfxsize=7.5cm
\centerline {\epsfbox[20 20 580 760]{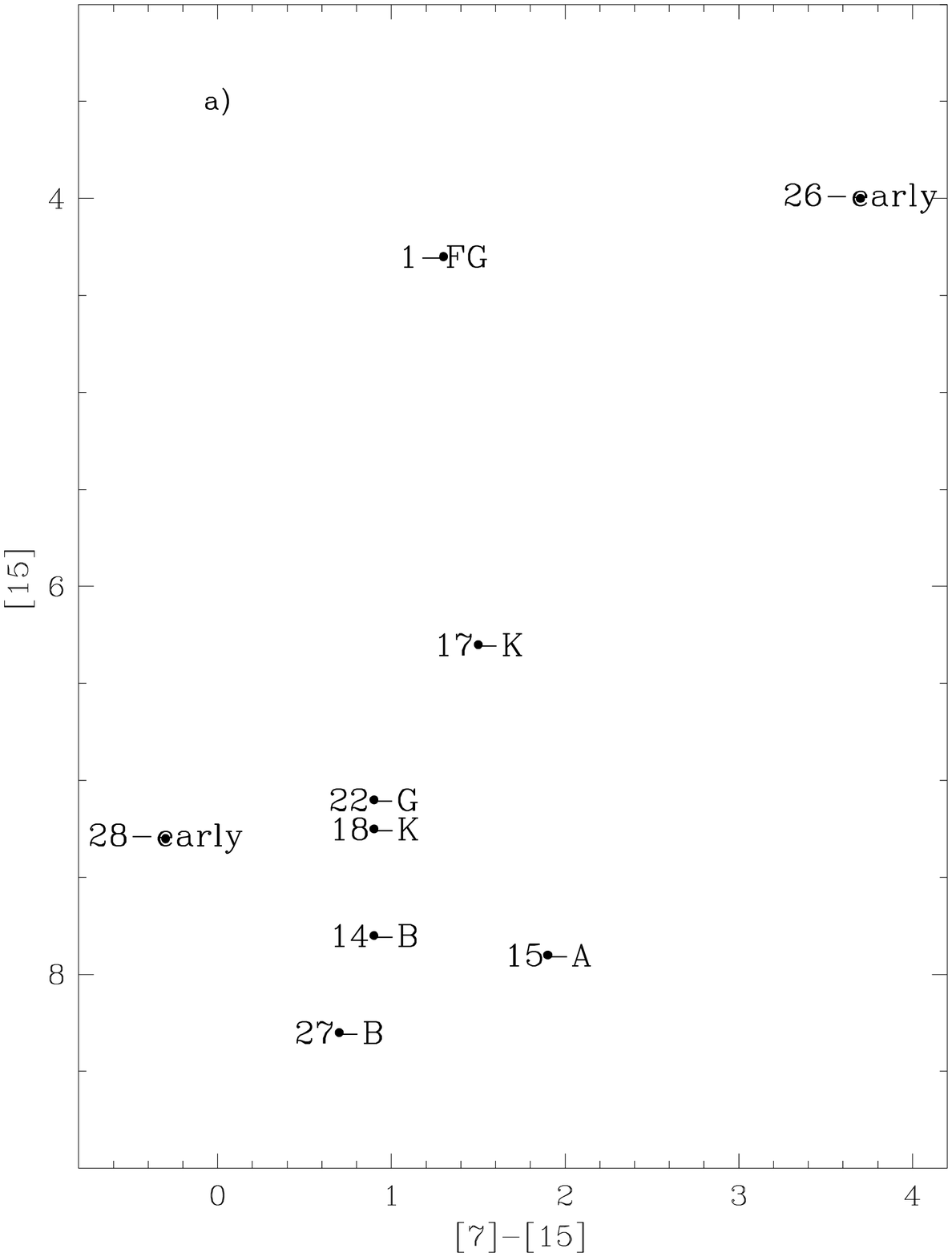} \epsfxsize=7.5cm
 \epsfbox[20 20 580 760]{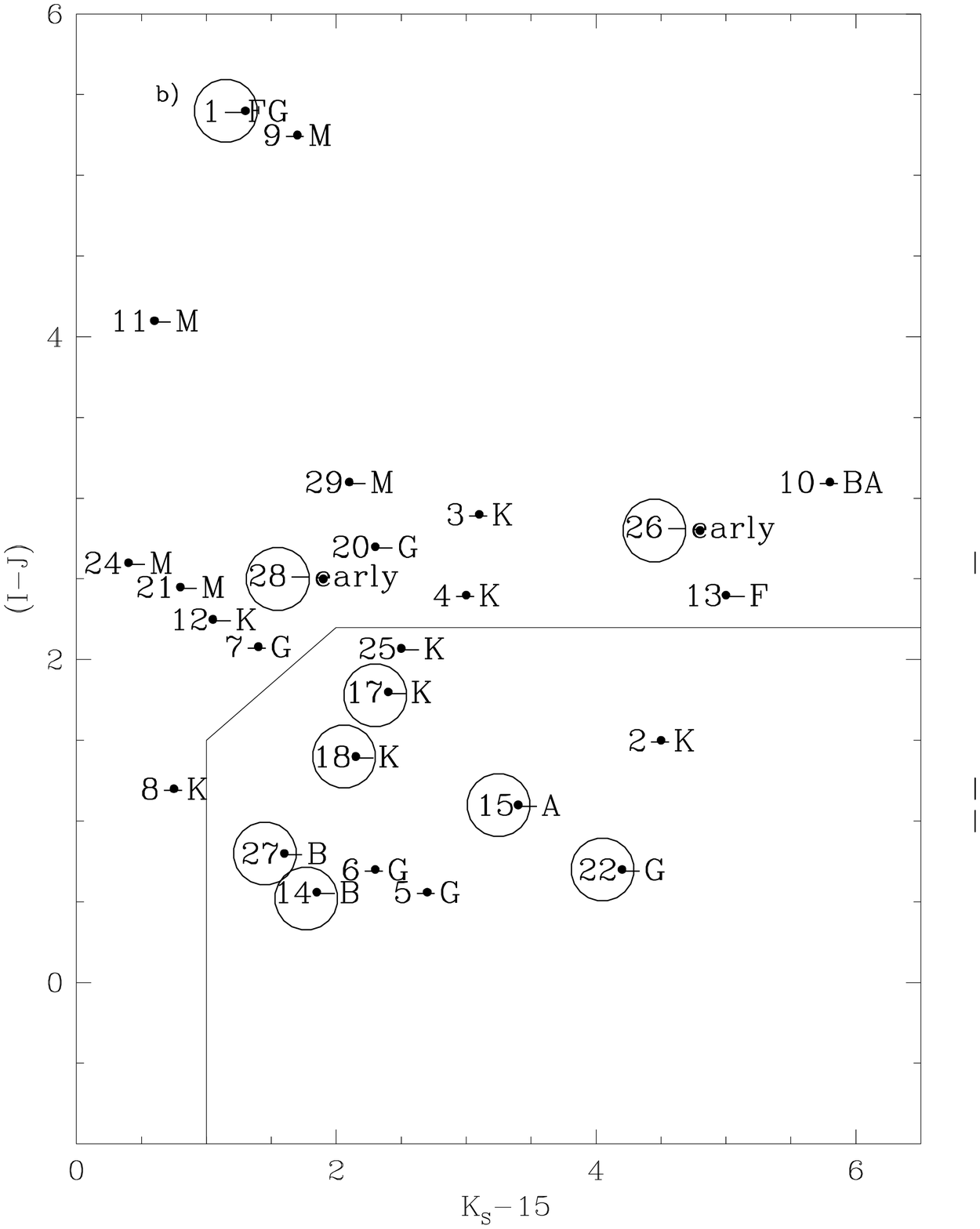}}
\caption{[15]/[7]-[15] diagram and $\rm (I-J)/K-[15]$ diagram of the ISOGAL
spectra. Besides the spectral type, the sequence number is indicated (see
Table 1). The ISOGAL sources marked with open circles are discussed in the main
text, while the others are presented in the Appendix A. The lines
indicate the approximate separation between early-type and late-type stars.}
\label{Figure1}
\end{figure*}
\begin{figure*}
\epsfxsize=7.5cm
\centerline {\epsfbox[20 20 580 760]{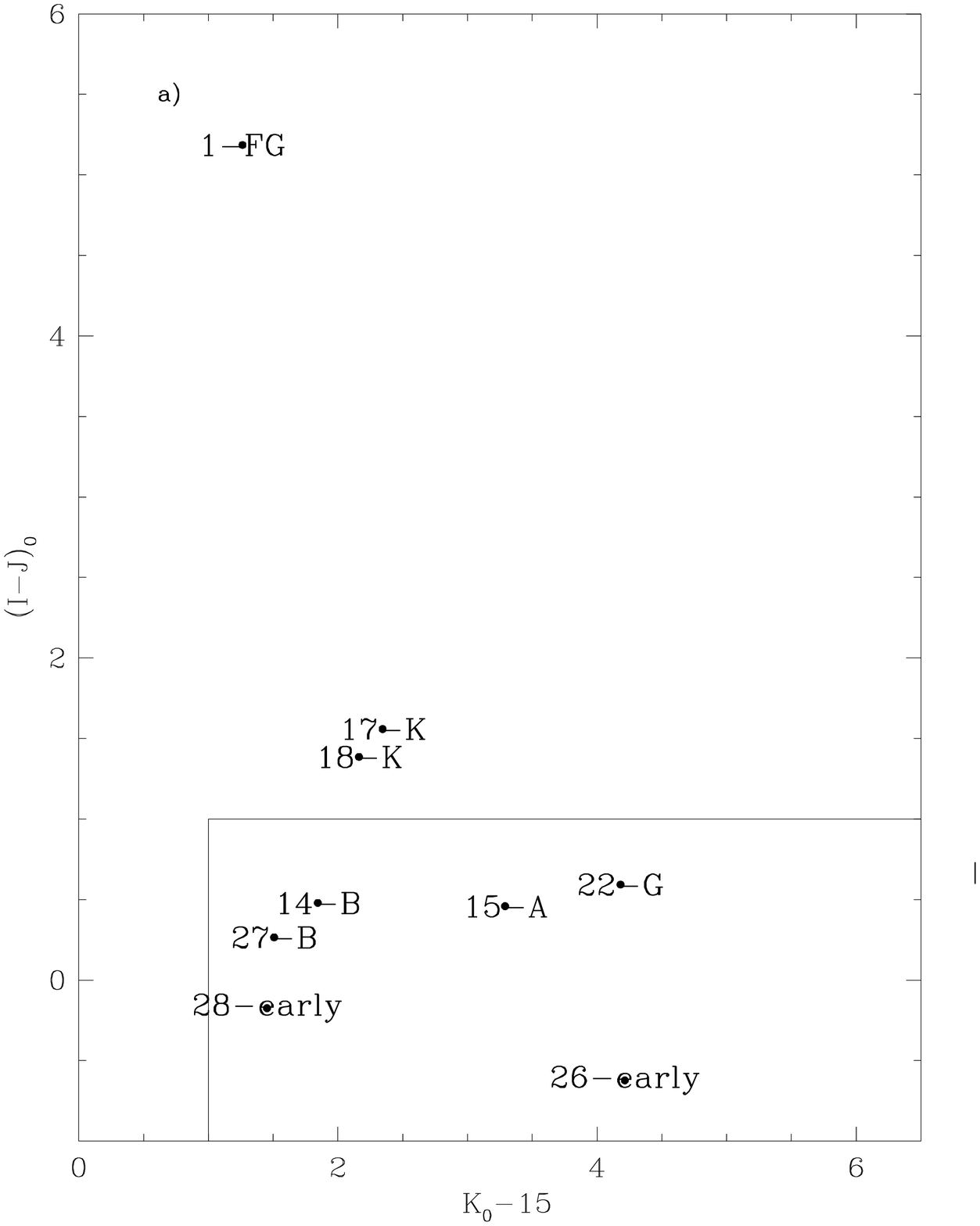} \epsfxsize=7.5cm
 \epsfbox[20 20 580 760]{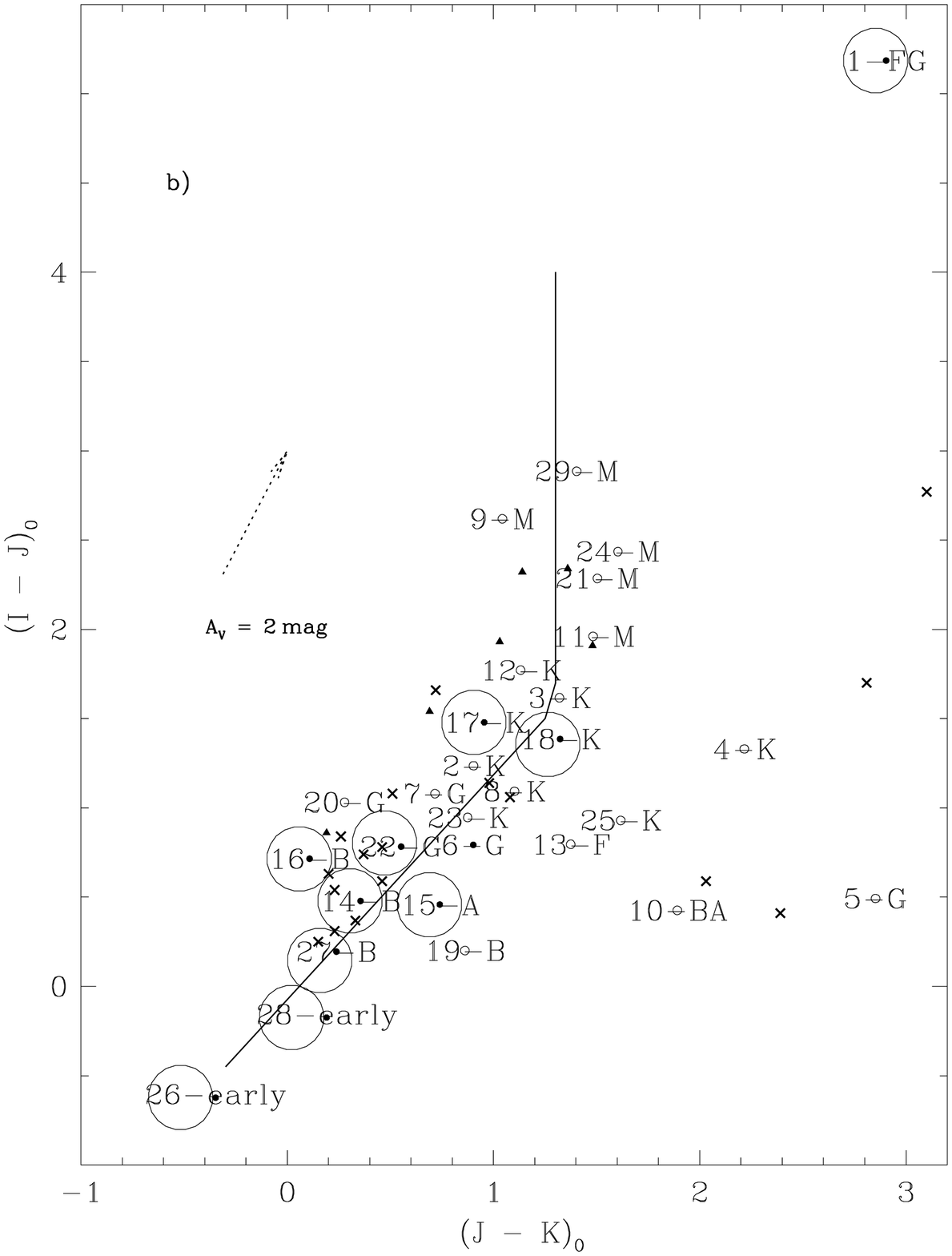}}
\caption{$\rm (I-J)_{0}/K_{0}-[15]$ diagram and $\rm (I-J)_{0}/(J-K)_{0}$ diagram of the ISOGAL
spectra. Besides the spectral type, the sequence number is indicated (see
Table 1). The straight lines in the left panel indicate the approximate separation between
 early-type and late-type stars. The lines in the right panel indicate
the temperature sequence for stars in the Pickles library, in the DENIS
system. In Figure~3b,  a sample of hot and cold post-AGB stars is superimposed, indicated
by crosses and triangles, respectively. The ISOGAL sources  marked with open circles are
discussed in the main text, while the others are presented in  the Appendix A
(see also Table~1). The extinction vector ($\rm A_{V} = 2\,mag$) is shown
by the dotted line. }
\label{Figure2}
\end{figure*}
\begin{figure}
\epsfxsize=8cm
\centerline {\epsfbox[30 110 680 530]{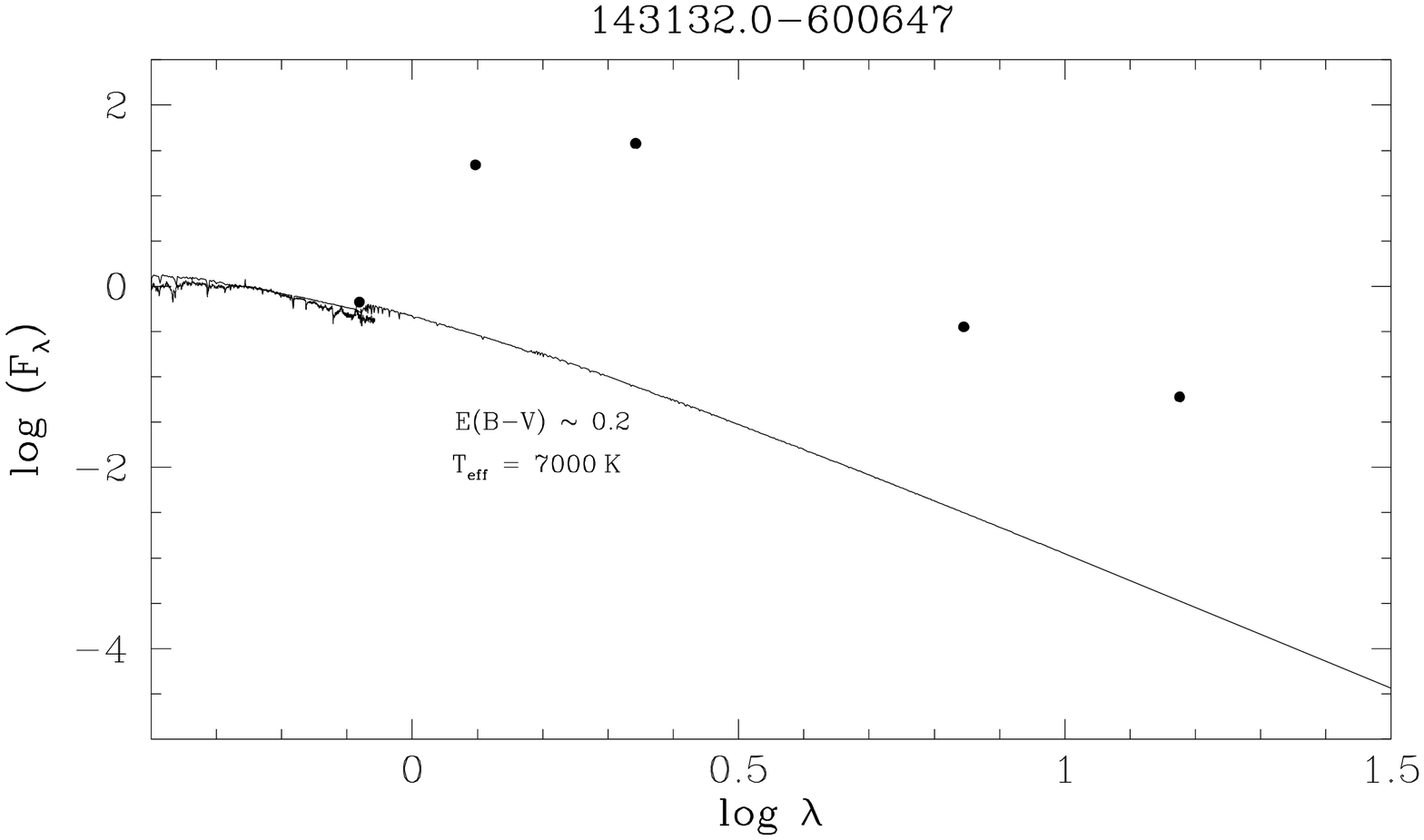}}
\caption{ Spectral energy distribution of object \#1 (F8-G0III). A
Kurucz model with $\rm T_{eff} = 7000\,K$ and a E(B-V) $\sim$ 0.2 gives
the best agreement to the spectrum. Note the strong near and mid-IR
excess of this object}
\label{SED}
\end{figure}

\begin{description}

\item \#1:   
In the colour-magnitude diagrams (see Figs.~\ref{Figure1} and ~\ref{Figure2}), this 
(F8-G0III) star, with narrow Balmer lines, shows a very strong near-IR excess and a moderate   mid-IR excess. 
The SED of this object (see Fig.~\ref{SED}) shows evidence for the presence
 of hot circumstellar dust with a temperature $\sim$1500K.
 Because our spectra are obtained with a slit spectrograph and the amount
of light going through the slit depends on the slit width and ``seeing'',
it is  hard to estimate the exact optical flux in units
of ergs/cm$^2$/sec. However, we find that the ratio of integrated IR flux
to integrated optical is about 50 which indicates that the
central star of object \#1 is significantly obscured.
 This high flux ratio between the IR and optical flux is similar to that
 of post-AGB stars with dusty disks
such as the Red Rectangle or Hen~401  (Parthasarathy et al.  \cite{Parthasarathy2001}). We find a small reddening of E(B-V) $\sim$ 0.2 indicating 
 that we see only the scattered starlight in the Optical.
High-resolution imaging in the IR may reveal
the dusty disk and bipolar geometry.
The large near-IR excess of this object is similar to that of some of the A and
F post-AGB stars with dusty disks and invisible binary companions
(van Winckel et al. \cite{Winckel95}). Radial velocity monitoring,
near-IR spectroscopy and photometric monitoring may enable us to probe further
 the evolutionary status of this star.


\item \#14:  
The object shows strong $\rm H_{\alpha}$ in emission while $\rm H_{\beta}$ is weak, apparently filled in by emission. A weak MgII absorption line at 4481\,\AA~ and diffuse interstellar bands (DIBs) at 5783\,\AA~ and 6288\,\AA~are visible. The colours (I--J)$_{0}$ and (J--K)$_{0}$ are consistent (Fig.~3b) with the spectral type. The moderate mid-IR excess could indicate a Vega-type star or a Herbig Be star. A second star is present $\sim$5$\arcsec$ away in DENIS and 2MASS with H=11.5 and $\rm K_{S} = 10.2$. However, its association with the ISOGAL source appears less probable because of its substantial projected distance. The possible extension of the 7 and 15\,$\mu$m source could, nevertheless, indicate a superposition of two objects.

\item \#15: 
This spectrum shows  a weak MgII line at 4481\,\AA~ and diffuse interstellar bands (DIBs) at 5783\,\AA~ and 6288\,\AA. There is a significant excess at 15\,$\mu$m which could indicate a Vega-type star. However, the position of the 15\,$\mu$m source at the edge of an extended emission region compromises the quality of its photometry. 

\item \#16: 
  The spectrum (B7) is similar to  \#15, with DIBs at 5783\,\AA~and 6288\,\AA. The $\rm H_{\alpha}$ line is broad and  partially filled in.
 The bump between 5000-6500\,\AA~ and the extended 15\,$\mu$m emission might indicate an object with extended red emission (see text). 

\item \#17:
In DENIS there is a second source with $\rm I = 13.9$, $\rm J = 12.8$ and $\rm K_{S} = 11.3$ at a large projected distance of 11\arcsec. However its near-IR magnitudes do not favour its association with the 7--15~$\mu$m source. The 
(I--J)$_0$ and (J--K)$_0$ colours of the main source are consistent. Therefore, we favour a K-giant with a  mid-IR excess.

\item \#18: 
 In DENIS there is a second source at a large projected distance of 
$\sim$ 10$\arcsec$ with $\rm I = 11.65$, $\rm J = 10.05$ and $\rm K_{S} = 9.27$. As for source \#17,
we favour a K-giant with mid-IR excess.

\item \#22: 
This object corresponds to a multiple system in DENIS/2MASS with projected distance of $\sim$ 2\arcsec.  We have not 
isolated the primary candidate because we suspect that the association of the visible/near-IR source and the ISOGAL source could  be spurious. However, the (I--J)$_0$ and (J--K)$_0$ colours follow the temperature sequence so we favour a G-star with mid-IR excess.

\item \#26:
Below 5000\,\AA~ the spectrum has poor S/N. HeI absorption lines at 5876\,\AA~ and 6678\,\AA~ may be present. It seems to be a heavily reddened early-type star. The 7\,$\mu$m image shows a nearly stellar object with a jet-like structure, while at 15\,$\mu$m it shows extended emission. While there is practically no excess at 7~$\mu$m, the excess at 15~$\mu$m is strong. This star is associated with IRAS 17485-2627 with high 60\,$\mu$m and 100\,$\mu$m fluxes
([12]$<$3.19, [25]=13.03, [60]=51.97, [100]=213.30), 
in a star-forming region (Codella et al. \cite{Codella95}). 
Although in J, H and K a second object has been extracted ($\rm J = 15.7$, $\rm H = 13.4$ and $\rm K_{S} = 12.0$) at $\sim$5\arcsec, it does not seem to be associated with the ISOGAL source. Consequently, we favour a young stellar object, with a relatively distant remnant accretion disk or cocoon, that produces the far-infrared excess.

\item \#27:
The spectrum (B3-4) is similar to that of \#14, with
$\rm H_{\alpha}$ in emission and DIBs at  5783\,\AA~and 6288\,\AA~. There is no doubt about the association and the mid-IR excess seems reasonably well-established. The bump between
5000-6000\,\AA~could be associated with extended red emission (see Section 5.3 )

\item \#28: 
The spectrum shows a strong reddening (E(B-V) $\sim$ 2.5) with many emission lines: [OI] line  at 5580\,\AA, $\rm H_{\alpha}$ ($\rm H_{\beta}$  is not present due to reddening), P~Cygni profile at 
6286\,\AA~ with an emission peak at 6300\,\AA,~[OI] emission line  at 6360\,\AA. There is a weak emission line on the
violet edge of the atmospheric feature at 6800\,\AA. The object seems to be a reddened early-type emission-line star. DIBs are present at 5783\,\AA~and 6288\,\AA. There is no doubt about the association of optical and IR objects. The mid-IR excess is strong. However, there is some inconsistency between the 7 and 15~$\mu$m magnitudes as the [7]-[15] colour is
 -0.3. The object lies in a region of extended emission at 7 and 15\,$\mu$m  and is within 3\,arcmin  of the open cluster, NGC~6531.

\end{description}

\section{Results and Discussion}

\subsection{Cross-identification between optical spectra and ISOGAL sources}

As mentioned in Sect.~4.3, we investigated the reliability of the
cross-identifications between the objects for which we tool optical 
spectra and the actual ISOGAL sources, in order to eliminate spurious 
associations. Therefore, for each individual source,
we examined by eye the DENIS/2MASS images in 
the four filters ({\it{I, J, H and $K_{S}$}}) as well as our ISOCAM 
images at 7 and 15\,$\mu$m.  In particular, we checked if there were
second, or multiple, sources close by which might be more plausible 
cross-identifications than our primary target stars.

For about 50\% of the sources, we did indeed find other nearby star
images.  Generally, these second components appear only
in the J, H or K bands, and are invisible in I.  We reject all such
sources from further discussion although we list them in Appendix A (see  the last column of Table 1).
Information about the cross-identifications is given
in the decription of the individual objects (Sect.~4.4 and Appendix A).
In the discussion below we will take into account only those sources 
where there is no doubt as to the association of the optical spectrum
with the ISOGAL counterpart.

\subsection{Colour-magnitude and colour-colour diagrams}

The infrared colour-magnitude and colour-colour diagrams are complementary to visible spectroscopy for analysing the nature of the sources. Four of the most interesting diagrams are displayed in Fig.~\ref{Figure1} and Fig.~\ref{Figure2}. Dereddened diagrams were produced with the estimated values of E(B-V) (see Table 1), except for the [15]/[7]--[15] diagram, which is somewhat insensitive to reddening. 

The [15]/[7]--[15] diagram shows that a few of the sources of Table 1 are among the brightest ISOGAL sources with a 15~$\mu$m flux up to 4~Jy, but that most of the other sources of this table are weak 15~$\mu$m  ISOGAL sources.
In order to ensure a reasonable level of reliability, completeness and
photometric accuracy, the ISOGAL data are limited to sources brighter
than 8.5\,mag (8\,mJy) at 15\,$\mu$m and 9.75\,mag (11\,mJy) at 7\,$\mu$m. 
We refer the reader to Omont et al. (1999) and Omont et al. (2002) for a detailed description of the ISOGAL catalog and its detection limits.
 However, the main interest of this diagram is to confirm that the sources discussed in section 4.4 (sources indicated by ``M'' in Table~1 except \#16 and \#28), have a significant excess at 15~$\mu$m with respect to 
7~$\mu$m. The 15~$\mu$m excess is exceptionally large in \#26, and is confirmed in the (I--J)$_0$/(K$_0$--15) diagram. The latter suggests that (I--J)$_0$ is an efficient discrimator between early spectral types and K/M types, except in the case of the peculiar F8-G0 giant \#1.

From the location in the $\rm (I-J)_{0}/(K_{0}-[15])$ diagram of the 
early-type stars with infrared excess, one can determine selection criteria to
seek the best candidates from the original DENIS-ISOGAL diagram, 
$\rm (I-J)/(K-[15])$. This colour-colour plane is a natural
one to look for non-AGB stars with circumstellar dust because of the SED
difference between AGB stars and hotter objects with a circumstellar shell.
Such objects are found in the domain shown in Fig.~2b. 
It has been designed with the dual objectives of not missing nearby 
early-type stars with a real mid-IR excess while limiting serious
contamination by distant AGB stars.

For near-infared colours, Fig.~3b displays the $\rm (I-J)_{0}/(J-K)_{0}$ diagram. The majority of the sources follow
a sequence which is well delineated by the stellar library  (temperature
range between 2500\,K to 35000\,K) of
Pickles (\cite{Pickles98}); such DENIS colours were calculated using
 the DENIS filter profiles and the mean atmospheric transmission at La Silla.
 The few objects on the right of this sequence could
be either cold pre-planetary nebulae with K excess (as defined by displaying
known examples in Fig.~3b), or the result of a spurious
association of  a red JK source with a faint blue I star.  
Coupled with visual inspection of the DENIS images, this
diagram indicates whether the photometry is affected by a nearby source.

\subsection{A and B spectral types}
Although our sample suffers from small number statistics,
the proportions of stars with various spectral types from B to K 
seem to be rather comparable to those of local Hipparcos stars
with 25\,$\mu$m IRAS excesses (Knauer et al. \cite{Knauer2001}).
	The majority of the stars retained have a mid-infrared excess associated with A or B spectral type. They probably belong to one of the important classes well studied by IRAS:

	- Young stars of intermediate mass with circumstellar dust. The most prominent are Herbig Ae-Be stars
(nebulous emission-line stars: see e.g.  Waters \& Waelkens \cite{Waters98}). However, other young A-B stars with similar infrared properties are also known without emission lines (see e.g. the analysis of IRAS sources in the molecular cloud L1641 by Strom et al. \cite{Strom89}).

	- B$_{[e]}$ with circumstellar emission lines (e.~g. Hummel et al. \cite{Hummel2001}).

	- Vega or $\beta$ Pic-like circumstellar debris disks. Although such disks are better detected at longer wavelength, the sensitivity of ISOGAL is such that stars similar to $\beta$ Pic should be detectable 
out to $\sim$1 kpc.

	- Post-AGB stars with infrared excess should be readily detected throughout the whole Galaxy. However, they are rare and mostly distant, so that our selection criteria will miss them in regions of high extinction. 

	Emission lines are found in  4 objects (H$\alpha$ emission is spectacular in \#10, but the association of the spectrum with the mid-infrared excess is dubious). The most probable explanation of such spectra would be Herbig Ae-Be stars. 

	It is also likely that \#26 is a young star, since it is located in the direction of a star-forming region with high extinction. The case of \#15 is not clear and it remains an interesting candidate for a $\beta$ Pic-type object at $\sim$1 kpc. High-resolution spectra are essential to
pursue the exact nature of such stars.

\subsection{A-B stars with a red excess}


	Two of the B stars are found to show a bump between
5000 and 6000\,$\rm \AA$~which could be due to extended red emission (ERE).
The most well-known object with ERE is the post-AGB star with the reflection nebula known as ``The Red Rectangle'' (HD44179)
(Schmidt et al. 1980).
 The red excess is clearly seen in the spectra of \#27 and \#16, and perhaps in \#14.  
ERE occurs in a wide variety of objects (Witt \& Boroson \cite{Witt90},
Gordon et al. \cite{Gordon98}), including  reflection nebulae such as
NGC 2023 and \mbox{NGC 7023}, while  Darbon et al. (\cite{Darbon2000}) found evidence of ERE in compact HII regions. 

The ERE in the spectrum of the Red Rectangle appears to be  due to several 
unidentified systems of emission bands (Schmidt et al. \cite{Schmidt80}), 
although $\rm CH^{+}$ has been identified in the blue (Balm \& Jura \cite{Balm93}). 
The ERE models of Seahra \& Duley (\cite{Seahra99}) predict a
spectrum with three peaks at 0.5, 0.7 and 1\,$\mu$m which are due to 
carbonaceous components of interstellar matter, while Gordon et al. (\cite{Gordon2000}) explain the ERE band by silicon nanoparticles. 
Witt et al. (\cite{Witt98})  found that silicon nanoparticles provide the
best match to the spectrum and to the photoluminescence efficiency required by the ERE.

The presence of  ERE in some of our B and A-type objects with $\rm H_{\alpha}$ emission
suggests that they may be also illuminate reflection
nebulae. High-resolution imaging in the
$\rm H_{\alpha}$ line and high-resolution spectroscopy may enable us to further understand
 the evolutionary stage, chemical composition, and ERE of these stars.

Two objects with ERE in our limited sample may be significant.
However,  to estimate the frequency of occurrence of ERE among
dusty A and B-stars requires a systematic spectral survey of a large
number of ISOGAL-DENIS sources, which is beyond the scope of the present
paper.



\subsection{FGK spectral types}


We also found that a few of our ISOGAL sources show the type of  mid-IR excess associated with
F,G and K-giants. Some of these could
be post-asymptotic giant branch stars (Oudmaijer et al. \cite{Oudmaijer92}, Pottasch \& Parthasarathy \cite{Pottasch88}), as suggested for FG-type  Hipparcos stars
(Knauer et al. \cite{Knauer2001}). In addition to five cases with dubious 
associations, we have two good candidates for such objects: \#22 and
the remarkable case of the F8-G0 giant \#1 (see Sect.~4.4).

 Zuckerman et al. (\cite{Zuckerman95}) and Plets et al. (\cite{Plets97}) found from an analysis of the IRAS point source
catalogue that about one percent of G and K-giants have warm circumstellar dust.
The presence of substantial dust near these stars is not easily explained and
various mechanisms have been proposed: binarity; evaporation of circumstellar
debris; or mass-loss on the first-ascent giant branch which could be related to
the overabundance of lithium observed in a fraction of these late-type giants
(de la Reza et al. \cite{Reza96},\cite{Reza97}, Jasniewicz et al. \cite{Jasniewicz99}, Fekel \& Watson \cite{Fekel98}). 
First-ascent giants are not expected to have significant mass-loss and
 circumstellar dust shells. It is, therefore, vital to study these objects with high spectral
resolution in order to understand their chemical composition and the origins of their
 circumstellar shells.

\section{Conclusions}


We have obtained  low-resolution spectra and IR data of 29 ISOGAL sources. 
After careful inspection of the DENIS/2MASS images and of the cross-identification
between ISOGAL and DENIS, about 50\% of the  objects have been rejected due to the influence
of nearby components. Six of the stars retained have 
A or B spectral type, probably mostly young stars.
Two of these show an extended red emission (ERE) indicating that they may be similar to the Red Rectangle
 or to reflection nebulae.
The spectral energy distribution of object \#1 suggests that it is 
probably a post-AGB stage in  a binary system where the dust is trapped in a  disc.

The $\rm (I-J)_{0}/(K_{0}-[15])$ and  (J--K)/(K--[15]) diagrams provide the most suitable tools to distinguish between early-type and late-type spectral types. The region of the $\rm (I-J)/K-[15]$ diagram
delineated in Fig.~2b is appropriate for a systematic search for nearby early-type objects
with infrared excess.

\appendix\section{Rejected Sources}

\begin{description}
\item \#2:
This source  is a very bright ISOGAL source. In $\rm K_{S}$ there is a second component 4$\arcsec$ away with $\rm K_{S} \sim 9.0$, but this object is invisible in I and J. This second object is more likely  to be associated with the ISOGAL source, while the IJK DENIS source corresponds to the observed visible spectrum.

\item \#3:   
This object shows a  strong feature at 6230\,\AA  where  TiO and perhaps CaOH
 are present. 
There is a second DENIS source (J = 14.8 and $\rm K_{S}$ = 12.1), $\sim$ 1\arcsec away, so that the association of the ISOGAL source with the observed visible spectrum is uncertain.
 
\item \#4:
The S/N of this spectrum is low but the FeI 5250\,\AA~ feature is strong.
At 7 and 15$\mu$m the object is embedded in extended emission.
The quality of DENIS images in J and K is poor but, in DENIS I, there is no sign
of a double component. However, the (J--K) and  (I--J) colours do not correspond to the temperature sequence, so that the association of the I source, its observed visible spectrum (K4), and the J--K--ISOGAL source is uncertain.

\item \#5:
The spectrum (G4) is very similar to that of  \#1, with a strong IR excess.
A double source in DENIS (projected distances of 3.2$\arcsec$ (primary candidate) and $\sim$ 2\arcsec), and inconsistent (I--J)/(J--K) colours make the association between the I source, its observed visible spectrum , and the ISOGAL source very 
questionable.

\item \#6:   
The spectrum (G4) is similar to that of  \#1.
At 7\,$\mu$m the source has been rejected from the new ISOGAL catalog. The ISOGAL association flag  makes the cross-identification between the DENIS and the weak 15~$\mu$m ISOGAL source questionable.

\item \#7:  
The S/N of this spectrum  is very weak  below 5000\,\AA,~due to strong reddening ($\rm E(B-V) \sim 1.0$). A second component in DENIS (projected distance of 2\arcsec) with $\rm J \sim 12.6$, $\rm K_{S} \sim 12.6$ and no detection in I makes any association with our ISOGAL source dubious.

\item \#8:
This weak ISOGAL object shows a double source in DENIS. The second component (projected distance of $\sim$ 2\arcsec) is not extracted by DENIS or 2MASS. 

\item \#9:  
Below 6500\,\AA~ the spectrum  has no 
 detectable flux, suggesting  strong reddening ($\rm E(B-V) \sim 2.6$).
The object has a strong near- and mid-IR excesses and is an AGB star
 with high mass-loss.

\item \#10:  
The spectrum shows very strong H$_{\alpha}$ emission and  emission lines at 4580\,\AA~
and 5070\,\AA. The stellar continuum is weak below 5200\,\AA;
 $\rm H_{\beta}$ is not visible perhaps due to the heavy reddening ($\rm E(B-V) \sim 2.5$).  There is a significant near-IR, and a large  mid-IR, excess. It could be an Ae-Be Herbi-type star. However, 
in the DENIS I image, a second very faint component ($\rm \sim 2\arcsec$ away) is seen but was not
extracted by DENIS. 

\item \#11:
This is a heavily reddened M giant (E(B-V) $\sim$ 2.0) with little near- or mid-IR excess. A double component (projected distance of $\rm \sim 3\arcsec$) appears with $\rm J \sim 13.5$, $\rm H \sim 11.9$ and $\rm K_{S} \sim 11.5$.

\item \#12:  
The spectrum (K5) is similar  to \#3. The TiO band  at 
6230\,\AA~is relatively strong. 
A second component at $\sim$ 5\arcsec) is visible in DENIS (not extracted by DENIS/2MASS). The
mid-IR excess is relatively weak and needs confirmation.

\item  \#13: 
The late-F visible spectrum is  similar to \#7.
However, there is a strong mid-IR excess and a significant near-IR excess. Such colours are known for some post-AGB stars. However, a companion is visible in the DENIS image (projected distance of $\rm \sim 2\arcsec$), which has been not extracted by DENIS and 2MASS. The inconsistent (J--K)$_0$ and  (I--J)$_0$ colours may, therefore, be due to a line-of-sight coincidence between an F-star and an AGB star.

\item \#19:  
The absorption spectrum shows a weak MgII line at 4481\,\AA~  and DIBs at 5783\,\AA~ and 6288\,\AA.
The strong near-IR and mid-IR excesses might indicate a Vega-type star, or even a post-AGB star. 
However, in DENIS a second red source is located $\sim$3$\arcsec$ away with $\rm K_{S} = 11.6$ and no I or J detection. The anomalous (I--J)$_0$ and (J--K)$_0$ colours could be contaminated by the second component. In addition, there is no confirmed 15~$\mu$m detection, and the photometry at 7\,$\mu$m could be affected by extended emission present in this region.
      
\item \#20:
This object is a multiple system in DENIS/2MASS (projected distance of $\rm \sim 2\arcsec$) whose other components are not extracted by DENIS/2MASS. The (I--J) and (J--K) colours indicate that the photometry might be affected by a second component. At 7 and 15\,$\mu$m, the region is very crowded.

\item \#21: 
 The spectrum shows an M star with a stellar continuum below 5000\,\AA~.

\item \#23:  
The continuum below 5000\,\AA~ is weak due to strong reddening
 ($\rm E(B-V) \sim 1.5$). The very faint point source at 7\,$\mu$m and 
the lack of confirmation of a 15$\mu$m conuterpart in the present ISOGAL catalog 
do not confirm the presence of a mid-IR excess.

\item \#24
This object is an M-giant with almost no detectable IR excess. 

\item \#25:
This is a double source in DENIS/2MASS (projected distance of $\rm \sim 2\arcsec$) which has not been extracted. The (I--J)
and (J--K) colours suggest a spurious identification between the ISOGAL and the DENIS sources.

\item  \#29:
This is a late M-giant with strong near- and mid-IR excesses. It is associated with  a known Mira Variable star (V1547 Sgr), with a period of $\sim$ 360 days.  

\end{description}

{\bf{Acknowledgements:}}

We would like to thank M.Gerbaldi and I.~S.Glass for reading the manuscript
and for the fruitful discussions.

MP thanks the IAP for their hospitality during visits financed 
 by the Indo-French IFCPAR collaboration project 1910-1.  

MS is supported by the Fonds zur F\"orderung der wissenschaftlichen
Forschung (FWF), Austria, under the project number J1971-PHY.

MC thanks NASA for its supporting this work through the ISO block grant to UC-Berkeley 
under contract 961501 through JPL.

The DENIS project is supported in France by the Institut National des
Sciences de l'Univers, the Education Ministry and the Centre National de la
Recherche Scientifique, in Germany by the State of Baden-W\"urtemberg, in
Spain by the DGICYT, in Italy by the Consiglio Nazionale delle Ricerche, in
Austria by the Fonds zur F\"orderung der wissenschaftlichen Forschung und
Bundesministerium f\"ur Wissenschaft und Forschung.

This publication makes use of data products from the Two Micron All Sky
Survey, which is a joint project of the University of Massachusetts and the
Infrared Processing and Analysis Center/California Institute of Technology,
funded by the National Aeronautics and Space Administration and the National
Science Foundation.

{}






\begin{figure*}
\epsfysize=4.5cm
\centerline {   \epsfbox[90 615 250 835]{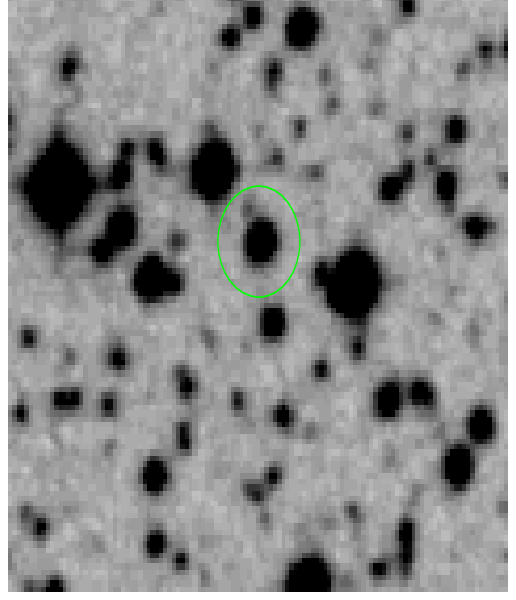} \epsfysize=4.5cm
   \epsfbox[90 615 250 835]{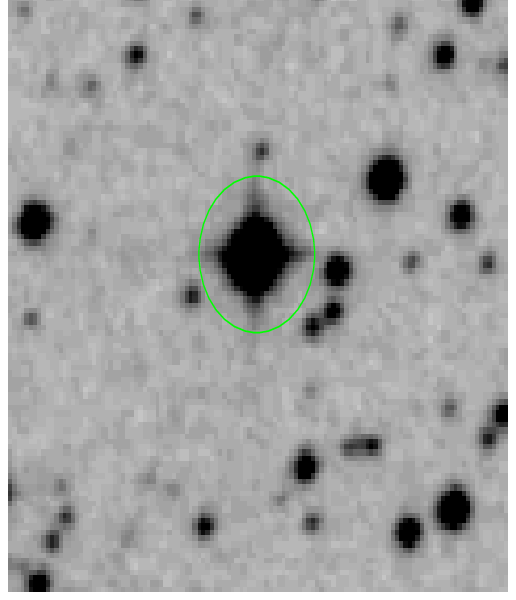}}

\epsfysize=4.5cm
 \centerline {\epsfbox[90 615 250 835]{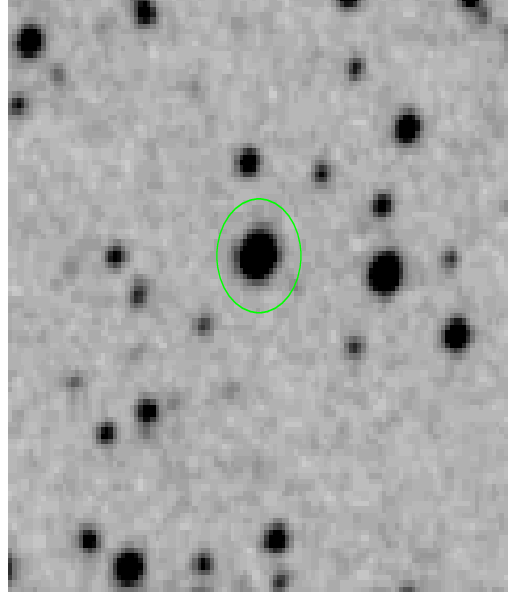} \epsfysize=4.5cm 
  \epsfbox[90 615 250 835]{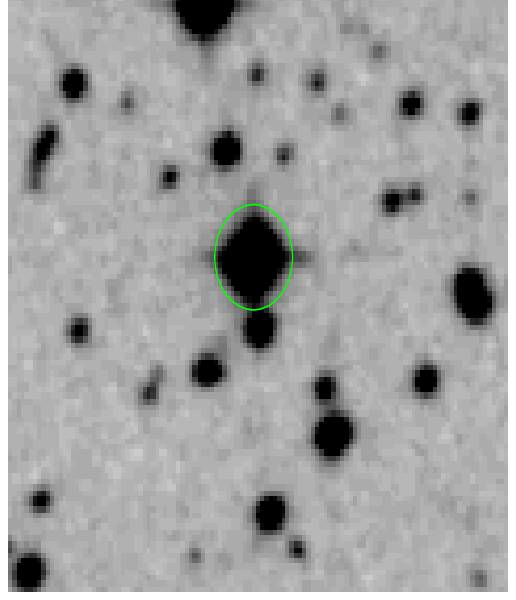}}

\epsfysize=4.5cm
\centerline {\epsfbox[90 615 250 835]{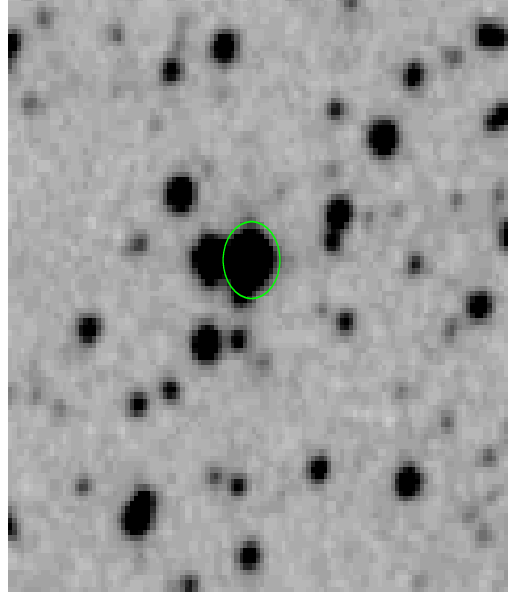} \epsfysize=4.5cm
 \epsfbox[90 615 250 835]{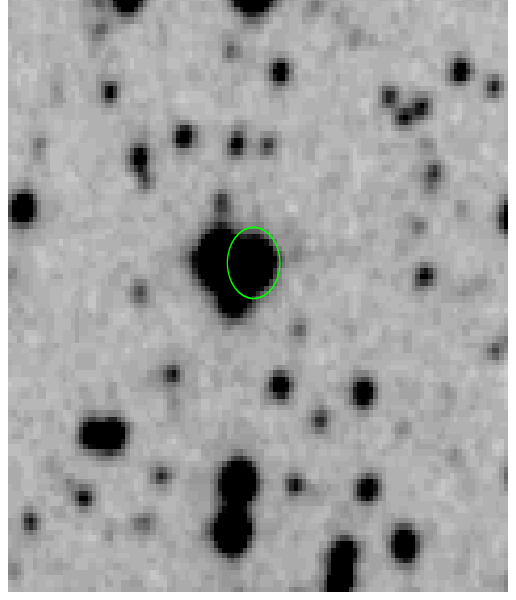}}

 \epsfysize=4.5cm
 \centerline  {\epsfbox[90 615 250 835]{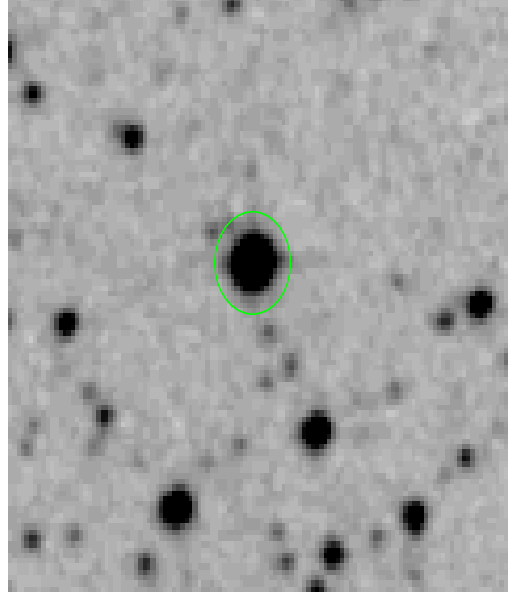} \epsfysize=4.5cm 
  \epsfbox[90 615 250 835]{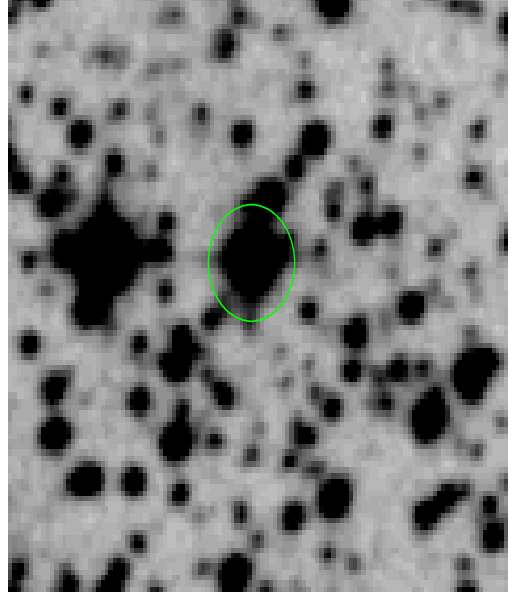}}

 \epsfysize=4.5cm
 \centerline {\epsfbox[90 615 250 835]{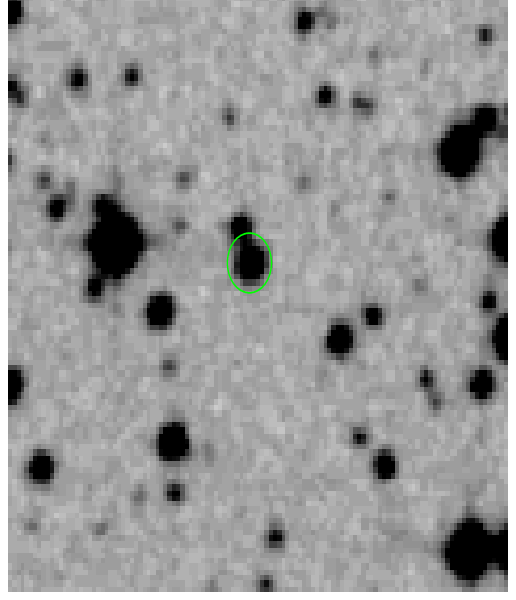} \epsfysize=4.5cm
 \epsfbox[90 615 250 835]{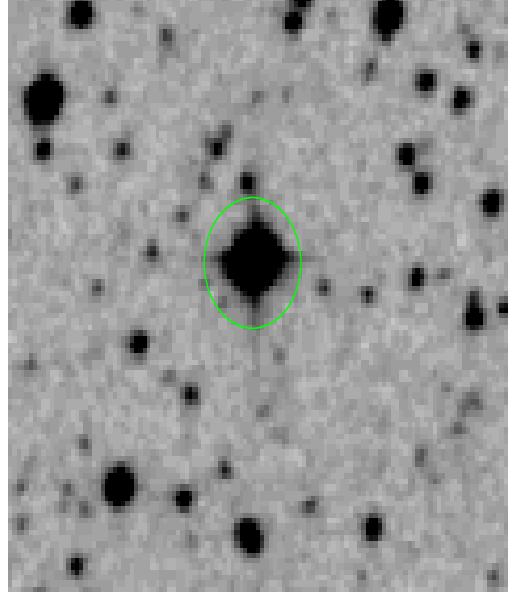}} 





\caption{ Finding charts in R of ISOGAL objects from the ESO online Digitized Sky 
Survey (DSS2).  The field of view is 2\arcmin x 2\arcmin around the center of the object.
North is up and east is left. See Table 1 for the correspondence with the coordinates.}

\end{figure*}

\end{document}